\documentclass[twoside,11pt]{article}

\usepackage{jmlr2e}

\usepackage[english]{babel}
\usepackage{natbib}
\usepackage{amsmath}
\usepackage{amssymb}  
\usepackage{amsfonts}
\usepackage{mathrsfs}
\usepackage{mathptmx} 
\usepackage{mathtools}
\usepackage{xcolor}

\newtheorem{thm}{Theorem}
\newtheorem{lem}{Lemma}
\newtheorem{prop}{Proposition}

\newtheorem{defn}{Definition}

\newtheorem{rem}{Remark}

\newtheorem{cond}{Condition}
\newtheorem{assm}{Assumption}

\usepackage{algorithm}
\usepackage{algpseudocode}
\usepackage[algo2e]{algorithm2e} 

\usepackage{multirow}
\usepackage{subcaption}
\usepackage{csquotes}

\newcommand*{\Scale}[2][4]{\scalebox{#1}{$#2$}}%

\jmlrheading{1}{2000}{1-48}{4/00}{10/00}{meila00a}{Majid Mazouchi,  Subramanya~Nageshrao, and Hamidreza~Modares}

\ShortHeadings{Finite-time Koopman Identifier}{Mazouchi, Nageshrao, and Modares}
\firstpageno{1}

\begin{document}

\title{Finite-time Koopman Identifier: A Unified Batch-online Learning Framework for Joint Learning of Koopman Structure and Parameters}

\author{\name Majid~Mazouchi \email mazouchi@msu.edu \\
       \addr Department of Mechanical Engineering\\
       Michigan State University\\
       East Lansing, MI, 48863, USA
       \AND
       \name Subramanya~Nageshrao \email snageshr@ford.com \\
       \addr  Ford Research and Innovation Center\\
       Ford Motor Company\\
       Palo Alto, CA 94304, USA
       \AND
       \name Hamidreza~Modares \email modaresh@msu.edu \\
       \addr Department of Mechanical Engineering\\
       Michigan State University\\
       East Lansing, MI, 48863, USA}

 \editor{}

\maketitle

\begin{abstract}

In this paper, a unified batch-online learning approach is introduced to learn a linear representation of nonlinear system dynamics using the Koopman operator. The presented system modeling approach leverages a novel incremental Koopman-based update law that regains a mini-collection of samples stored in a memory to minimize not only the instantaneous Koopman operator's identification errors but also the identification errors for the collection of retrieved samples. Discontinuous modifications of gradient flows are presented for the online update law to assure finite-time convergence under easy-to-verify conditions defined on the batch of data. Therefore, this unified online-batch framework allows joint sample- and time-domain analysis to converge the Koopman operator's parameters. More specifically, it is shown that if the collected mini-batch of samples guarantees a rank condition, then finite-time guarantee in the time domain can be certified, and the settling time depends on the quality of collected samples being reused in the update law. Moreover, the efficiency of the proposed Koopman-based update law is further analyzed by showing that the identification regret in continuous time grows sub-linearly with time. Furthermore, to avoid learning corrupted dynamics due to the selection of an inappropriate set of Koopman observables, a higher-layer meta-learner employs a discrete Bayesian optimization algorithm to obtain the best library of observable functions for the operator. Since finite-time convergence of the Koopman model for each set of observables is guaranteed under a rank condition on stored data, the fitness of each set of observables can be obtained based on the identification error on the stored samples in the proposed framework and even without implementing any controller based on the learned system. Finally, to confirm the effectiveness of the proposed scheme, two simulation examples are presented.
\end{abstract}


\section{Introduction}

Koopman operator theory provides an elegant framework to construct (infinite-dimensional) linear representations for nonlinear dynamical systems from input-output data and, as a result, a solid framework to use established model-based linear control methods. More specifically, Koopman operator theory relies on a nonlinear transformation to lift the original state-space to a space of functions, referred to as observable functions, under which their flow along trajectories of the system is described by a linear map with generally an infinite-dimensional state space. Finite-dimensional approximations of the Koopman operator from data has been widely considered  \citep{budivsic2012applied,bakker2019koopman}, mainly using Dynamic Mode Decomposition (DMD) \citep{tu2013dynamic} and Extended Dynamic Decomposition (EDMD) \citep{williams2015data}.  These approaches are mainly presented for discrete-time (DT) systems. Continuous-time (CT) formulations for system dynamics, however, can be methodologically preferred over DT formulations for control of many systems, such as safety-critical systems. For example, control-theoretic analyses such as Nagumo's theorem (which is fundamental for characterizing the forward invariance of safe sets) are not applicable for general DT systems \citep{maghenem2018barrier}. Despite its advantages, CT Koopman representation is scarce \citep{kaiser2021data}.

Koopman operator-based identification techniques mainly use a large batch of data samples as full training data set to learn offline the Koopman model. On the one hand, since the size of the required batch of data set is generally huge for high-dimensional Koopman-based representations, training on the full data at once can incur a high cost of computing resources. 
Moreover, a majority of practical control systems are not able to wait until gathering a large and rich dataset and learning the accurate model to make decisions, and they need to make decisions by learning the model on the fly by using only available data samples which incrementally become available over time \citep{ayoobi2021argumentation}.
 On the other hand, even though online or incremental learning allows fast and computationally cheap (at least sub-linear time and space complexity) learning algorithms, it can only provide generalization guarantees (i.e., learning the exact system parameters) if a persistency of excitation condition on the richness of data is satisfied, which is hard to guarantee when samples are collected incrementally over time. 

Another factor that has a huge impact on the performance of the Koopman-based identification, besides its learning rule, is the selection of observable functions. EDMD can embody Koopman observables (i.e., eigenfunctions) that are nonlinear functions of the state (Generally sets of radial basis functions \citep{williams2016extending,korda2018linear} and polynomials \citep{williams2015data,proctor2018generalizing}). Even though a large size of observables might lead to better accuracy, it is desired to learn reduced-order Koopman-based models that are amenable to control-system approaches \citep{ williams2015data, kutz2016koopman}. Moreover, some eigenfunctions can be distorted when projected into a finite-dimensional measurement subspace, causing undesirable performance degradation in the learning process. Instead of choosing a dictionary of observable functions a priori, another approach is to use neural networks to learn the observable functions from the data \citep{ yeung2019learning,li2017extended}. These methods, however, typically separate learning the Koopman observables and its model parameters and are based on the availability of a huge batch of i.i.d samples. 

In sharp contrast to the existing offline Koopman operator identifier, a novel data-driven bilevel learning algorithm is presented to jointly learn the efficient set of observables and the corresponding Koopman parameters. The proposed learning scheme is composed of two layers: a lower layer parameter learning and a higher layer structure learning. In the lower layer (so-called base learner), a unified batch-online learning framework is presented for learning the system Koopman operators' parameters through an experience-replay gradient descent (ERGD) \citep{modares2013adaptive,chowdhary2013concurrent,yang2014near,liu2014experience,chowdhary2010concurrent,kamalapurkar2017concurrent,vamvoudakis2015asymptotically,zhao2015experience,zhang2016event,cao2017robust,jha2019initial,parikh2019integral,tatari2017distributed,tatari2018optimal,walters2018online,yang2019adaptive,jiang2019path,yang2020safe,vahidi2020memory} update law that not only minimizes the current Koopman operator's identification errors online but also the identification errors for a batch of past samples collected in a history stack. To achieve finite-time convergence guarantee  \citep{lu2016note,he2017finite,Romero2021TimevaryingCO,xia2019finite,song2019neural,romero2020finite,zhao2020finite,Romero2020RobustTC,Tatari2021FinitetimeIO} of the presented Koopman learning approach with a prescribed settling time, a discontinuous ERGD update law is presented and Filippov differential inclusions along with finite-time Lyapunov stability theory are leveraged to analyze these discontinuous flows and the performance of the Koopman learned model. It is shown that the settling time and generalization capability of the learned model depend on the maximum eigenvalue of the matrix formed by the mini-batch of collected data that are being continually reused during online learning. The efficiency of the proposed Koopman-based update law is further analyzed by showing that the identification regret in continuous time grows sub-linearly with time.

In the higher layer (so-called meta-learner), a discrete Bayesian optimization algorithm \citep{brochu2010tutorial} is used to learn the best library of observable functions. More specifically, adapting the lifted state (library of observable functions) of the base learner is performed in a higher layer to develop a model that is of high fidelity but also has as low dimensionality as possible. Since finite-time convergence of the Koopman model for each set of observables is guaranteed under a rank condition on stored data, the fitness of each set of observables can be obtained based on the identification error on the stored samples in the proposed framework and even without implementing any controller based on the learned system.  Finally, two simulation examples are given to verify the effectiveness of the proposed scheme.

%
%

The rest of this paper is organized as follows: Some preliminaries are provided in Section 2. The problem formulation is provided in Section 3. The main theoretical results are provided and proved in Section 4. Section 5 presents some simulation results to verify the performance of the proposed control scheme. Finally, Section 6 provides the conclusion for the paper.

{\bf Notation.} Throughout the paper, $\mathbb{R}$  and   ${\mathbb{R}}^+$  represent the set of real numbers and positive real numbers, respectively. ${\mathbb{R}}^n$  and ${\mathbb{R}}^{n \times m}$  denote $n$-dimensional Euclidean space and $(n \times m)$-dimensional real matrices space. $I$ denotes the identity matrix of the proper dimension.  $\parallel .\parallel$  denotes the Euclidean norm of a vector or the induced $2$-norm of a matrix.
  We use ${^{\top}}$ to denote the transpose operator.
   $A > 0$  denotes that the matrix $A$ is positive definite.
    We use ${\lambda _{min}}(A)$ (resp. ${\lambda _{\max }}(A)$) to denote the minimum (resp. maximum) eigenvalues of the square matrix $A$. The Kronecker product of two matrices  $A$  and  $B$ is denoted as $A \otimes B$. For a function $f$, $f \ge 0$ denotes that $f$ is positive semi-definite. ${\bf K}:{S_1} \mathbin{\lower.3ex\hbox{$\buildrel\textstyle\rightarrow\over
{\smash{\rightarrow}\vphantom{_{\vbox to.5ex{\vss}}}}$}} {S_2}$ denotes a set-valued map  $\bf K$ for given sets ${S_1}$  and  ${S_2}$ such that $s \in {S_1} \Rightarrow {\bf K}(s) \subseteq {S_2}$.

\section{Preliminaries}
Some basic preliminaries are briefly introduced in this section. 
Consider a vector differential equation described in the form of
\begin{align}
\begin{array}{l}
\dot x(t) = {\cal I}(x(t)) \\
x(0) = {x_0}
\end{array} \label{eq:1}
\end{align}
where  ${\cal I}:{\mathbb{R} ^n} \mapsto {\mathbb{R} ^n}$  may be discontinuous, but Lebesgue measurable inside an open region ${\cal{R}} \subset {\mathbb{R} ^n}$, ${\cal{R}}  \ne \emptyset $,  and essentially locally uniformly bounded (i.e., ${\cal I}$  is bounded a.e. (almost everywhere) inside a bounded neighborhood of the point  $ x \in {\mathbb{R} ^n}$, $\forall x$). 
Note that the discontinuous system \eqref{eq:1} can be seen as a Filippov differential inclusion, and in this sense, the following definition formally describes a solution of \eqref{eq:1}.

{\begin{defn}\label{defn:1} \citep{filippov2013differential,paden1987calculus}
A Filippov solution of the vector differential equation \eqref{eq:1} on an interval $[0,\tau )$ is defined to be an absolutely continuous function  $x:[0,\tau ) \mapsto {\mathbb{R} ^n}$ with  $0<\tau \le \infty $ such that $x(0) = {x_0}$ and 
 \begin{align}
\dot x(t) \in {\bf K}[{\cal I}](x(t)), \label{eq:2}
\end{align}
holds for almost all  $t \in [0,\tau )$, where the Filippov set-valued map ${\bf K}[{\cal I}]:{\mathbb{R} ^n} \rightrightarrows {\mathbb{R} ^n}$ is  described by 
\begin{align}
{\bf K}[{\cal I}](x) \equiv \bigcap\limits_{\delta  > 0} ~ {\bigcap\limits_{S:~\mu S = 0} {\overline {{\mathop{\rm co}\nolimits} \,} } } \{{\cal I}\left( {{\cal B}(x,\delta ) 	\setminus S} \right)\}, \label{eq:3}
\end{align}
where $\overline {co} $ denotes the convex closure, ${\cal B}(x,\delta )$ denotes the open  ball of radius $\delta $ with the center $x$,  $\mu $ denotes the Lebesgue measure,
  , and $\bigcap\limits_{S:~\mu S = 0}$ denotes the intersection over all sets $S$ where $S$ runs over all zero-measure sets of $\mathbb{R}^n $.
Moreover, the Filippov solution $x$ is called maximal, if it cannot be extended \citep{Corts2008DiscontinuousDS}. 
\end{defn}}

{
	\begin{defn}\label{defn:20}
		\citep{thieme2019multiflows} A set-valued function ${\bf K}[{\cal I}]:{\mathbb{R} ^n} \rightrightarrows {\mathbb{R} ^n}$ is called upper semi-continuous at the point $x$ if, for any $\epsilon>0$, there exists some $\delta>0$ such that ${\bf K}[{\cal I}]\left({\cal B}(x,\delta)\right)$ is a subset of an open $\epsilon$-neighborhood of ${\bf K}[{\cal I}](x)$. If ${\bf K}[{\cal I}]$ is upper semi-continuous at each point $x \in \mathbb{R}^n$, then it is said to be upper semi-continuous.
\end{defn}}

\begin{cond}\label{cond:1}
${\bf K}[{\cal I}]:{\mathbb{R} ^n} \rightrightarrows {\mathbb{R} ^n}$ is upper semi-continuous, and has compact, nonempty, and convex values.
\end{cond}

{
\begin{lem}\label{lem:1}
	\citep{paden1987calculus}
Consider Filippov differential inclusion \eqref{eq:2}. 
There exists a zero-level set ${{\cal N}_{\cal I}} $, i.e., $\mu {{\cal N}_{\cal I}} = 0$, such that  \eqref{eq:3} can be computed as 
\begin{align}
{\bf K}[{\cal I}](x) ={\overline {\mathop{\rm co}\nolimits}}  \left\{ {\mathop {\lim }\limits {\cal I}\left( {{x_i}} \right)| \, {x_i} \to x}, \, {x_i} \notin { S \cup {\cal N}_{\cal I}} \right\}, \label{eq:4}
\end{align} 
where $S $ is any Lebesgue-zero measure set, i.e., $\mu S = 0$.
 In particular, ${\bf K}[{\cal I}](x) = \{ {\cal I}(x)\}$ provided that ${\cal I}$ be continuous at the fixed point $x$.
\end{lem}}

Now, let us introduce generalized derivatives and gradients to deal with non-smooth Lyapunov functions.

\begin{defn}\label{defn:2}
Let $V( \cdot ):{\mathbb{R} ^n} \mapsto \mathbb{R} $ be a locally Lipschitz function. Then, the Clarke upper generalized derivative of  $V(.)$ at $x$ in the direction of $v$ (also called generalized directional derivative) is defined as
\begin{align}
{V^\circ }(x;v): = \mathop {\lim \sup }\limits_{{x^\prime } \to x\,\,t \to {0^ + }} \frac{{V\left( {{x^\prime } + tv} \right) - V\left( {{x^\prime }} \right)}}{t}. \label{eq:5}
\end{align}
\end{defn}

\begin{defn}\label{defn:3}
$V( \cdot ):{\mathbb{R} ^n} \mapsto \mathbb{R} $ is called regular if for all $v$:\\
1) the usual one-sided directional derivative  $V( \cdot )$
\begin{align}
{V^\prime }(x;v): = \mathop {\lim }\limits_{t \to 0} \frac{{V(x + tv) - V(x)}}{t}, \label{eq:6}
\end{align} 
exists for all $v$;\\
2) ${V^\circ }(x;v) = {V^\prime }(x;v)$.
\end{defn}

\begin{defn}\label{defn:4}
The Clarke generalized gradient of $V( \cdot )$  at $x$  is the set 
 \begin{align}
\partial V(x) = {\mathop{\rm co}\nolimits} \left\{ {\mathop {\lim }\limits_{i \to  + \infty } \nabla V\left( {{x_i}} \right):{x_i} \to x,\;\;\,{x_i} \notin S \cup {{\cal N}_V}} \right\}, \label{eq:7}
\end{align} 
where $ {\mathrm{co}}$ denotes convex hull and  $S \subset {\mathbb{R} ^n}$ is any zero-measure set and ${{\cal N}_V}$  the zero-measure set over which $V$ is not differentiable.
\end{defn}

\begin{lem}\label{lem:2}
\citep{bacciotti1999stability}
Let $V( \cdot )$ be a locally Lipschitz and regular function. Then, \\
1)   \begin{align}
\partial V(x) = \left\{ {\xi  \in {\mathbb{R} ^n}:{V^o}(x,v) \ge \xi  \cdot v\;\;\,\forall v \in {\mathbb{R} ^n}} \right\}; \label{eq:8} 
\end{align}  
2)   \begin{align}
{V^{\circ} }(x;v) = \max \{ \langle \xi ,v\rangle  = \xi .v\mid \xi  \in \partial V(x)\};  \label{eq:9}
\end{align} 
3)  \begin{align}
{\bf K}[\nabla V](x) = \partial V(x). \label{eq:10}
\end{align}   
\end{lem}

\begin{cond}\label{cond:2}
 {$V( \cdot )$ is a locally Lipscthiz continuous and regular Lyapunov function}
  \footnote{{In other words, it is a scalar function that is continuous, has continuous first derivatives, is strictly positive, except the origin, and has a non-positive time derivative $\dot{V}$.}}
  {for the system \eqref{eq:1}.}
\end{cond}

{
\begin{defn}\label{defn:21} 
	\citep{bacciotti1999stability}
Given a locally Lipschitz function $V(.): \mathbb{R}^n \mapsto \mathbb{R}$ and a set-valued map ${\bf K}[{\cal I}](x)$,
the set-valued time derivative of $V$ with respect to the differential inclusion \eqref{eq:2} is defined as
\begin{align}
	\dot V(x) \triangleq \left\{a \in \mathbb{R}: \exists v \in {\bf K}[{\cal I}](x) \text { s.t. } a=p^{\top} v, \, \forall p \in \partial V(x)\right\},
\end{align}
for each $x \in \mathbb{R}^n$.
\end{defn}
 }



The following proposition provides a sufficient condition for finite-time convergence of differential inclusions.

\begin{prop}\label{prop:1}  \citep{romero2020finite} 
Consider the system \eqref{eq:1} and let Condition 1 be satisfied for a set-valued map $ {\bf K}:{\mathbb{R} ^n} \mathbin{\lower.3ex\hbox{$\buildrel\textstyle\rightarrow\over
{\smash{\rightarrow}\vphantom{_{\vbox to.5ex{\vss}}}}$}} {\mathbb{R} ^n}$.
 Let  ${\cal R} \subseteq {\mathbb{R} ^n}$  be a positively invariant and open neighborhood of ${x^*}$ and $V:{\cal R} \mapsto \mathbb{R}$ be positive definite 
 \footnote{{In other words, there exists some open neighborhood ${\cal R}$ of $x^{\star}$ such that $V$ is defined in ${\cal R}$ and satisfies $V\left(x^{\star}\right)=0$ and $V(x)>0$ for every $x \in {\cal R} \backslash\left\{x^{\star}\right\}$.}} 
 satisfying Condition 2.
 Consider the constants $c_1 > 0$  and  $c_2  < 1$ exist such that
\begin{align}
\sup \dot V(x) \le  - c_1 V{(x)^{c_2} }, \label{eq:11}
\end{align}
a.e. in $x \in {\cal R}$. 
Then, \eqref{eq:1} converges to  ${x^*}$ within finite time and a settling-time which is upper bounded by
\begin{align}
{t^ \star } \le \frac{{V{{\left( {{x(0)}} \right)}^{1 - c_2 }}}}{{c_1 (1 - c_2 )}}. \label{eq:12}
\end{align} 
\end{prop}

\section{Problem Formulation}
The finite-time identification problem is formulated in this section for a continuous-time nonlinear system with dynamics given by
\begin{align}
\dot x = F(x) + G(x)u(t), \,\,\, F(0) = 0, \label{eq:13}
\end{align}
where $x \in {\mathbb{R} ^n}$  and  $u(t) \in {\mathbb{R} ^m}$   denote the state vector and the control input, respectively. Furthermore,  $F(x) \in {\mathbb{R} ^n}$ and $G(x) \in {\mathbb{R} ^{n \times m}}$   denote the drift dynamics and the input dynamics of the system, respectively.

\begin{assm}\label{Assm:1}
{The functions $F(x)$ and $G(x)$ are smooth. System states are available for measurement. However, the derivatives of the states of system may not available for measurement.}
\end{assm}

{The key step in obtaining a linear structured version of a nonlinear dynamical system \eqref{eq:13} is a lifting of the state-space \citep{mauroy2019koopman,drmavc2021identification} to a higher-dimensional space, where its evolution is approximately linear.
Let us consider real-valued observables elements 
$\xi_i$, which are elements of an infinite-dimensional Hilbert space. Generally, the Hilbert space is provided by the Lebasque-square integrable functions. 
In this study, it is our aim to identify the nonlinear vector fields ${ F}(.)$ and ${ G}(.)$ based on stream of data generated by the control dynamical system and find a continuous-time dynamical system of the form \citep{huang2020data,brunton2016koopman,korda2018linear}
\begin{align}
	\frac{d}{{dt}}\xi ({ t}) = A\xi ({ t}) +  B \Psi (u(t)), \label{eq:15}
\end{align}
where the term $B \Psi (.)$  captures how the observable  $\xi (x)$ is modified by the control policy $u(.)$, e.g., $\Psi (u(t))$ is chosen as $\xi ({x})u(t)$ in \citep{korda2018linear}. }

{
	\begin{defn}\label{defn:5}
	The finite set of all available observables, denoted by ${\cal C}$, is defined as ${\cal C}: = \big\{ {\xi _1}(x),{\xi _2}(x),$ $ \ldots,{\xi _N}(x) \big\}$. A library set $ {\cal L}(\theta )$,  where $ {\cal L}(\theta ) \subseteq {\cal C}$, is defined as ${\cal L}(\theta ) : = \left\{ {{\xi _k}:k \in \theta, {\xi _k} \in {\cal C}} \right\}$ where  $(\theta, \le ) \subseteq \{ 1,...,N \} $ is a non-empty finite partially ordered set with ${n_{{\xi _\theta }}} := \left| \theta  \right|$  distinct elements. 
		The library vector corresponding to $\theta$, i.e., ${\cal L}(\theta ) $, is defined as ${\xi _\theta }(t):= [ {\xi _{\theta (1)}}(x(t)),$ $ \cdots, {\xi _{\theta({n_{\xi _\theta }})}}(x(t)) ]^{\top}$ .
\end{defn}}

{A library vector ${\xi _\theta }(t)$ corresponding to the ordered set of  $\theta$, i.e., ${\cal L}(\theta )$, which is a subset of the finite set of all available observables ${\cal C}$, will be used in the sequel. It is worth noting that the library vector ${\xi _\theta }(t)$ will be selected by the meta-learner layer (as detailed in Subsection 4.3) to lift the system from a state-space to function space of observables with the benefit of reducing the dimension of the Koopman operator and avoiding selecting corrupted dynamics. Note that the dimension of the library vector ${\xi _\theta }(t)$ is the same as the size of set ${\cal L}(\theta )$, i.e.,  ${\xi _\theta }(t) \in {\mathbb{R} ^{{n_{{\xi _\theta }}}}}$. In the rest of the paper, for simplicity of notation, we have used ordered set $\theta $ instead of the library of observables ${\cal L}(\theta )$. }

{Now, the infinite-dimensional Koopman operator is approximated with a finite-dimensional matrix $A_\theta ^*$  and  $B_\theta ^*$, and the linear representation continuous-time dynamics are given by}
\begin{align}
\frac{d}{{dt}}{\xi _\theta }(t) &= A_\theta ^*{\xi _\theta }(t) + B_\theta ^*\Psi (u(t)) + {{\cal P}_\theta }({{\xi _\theta }(t) }),
\label{eq:16}
\end{align}
where $A_\theta ^* \in {\mathbb{R} ^{{n_{{\xi _\theta }}} \times {n_{{\xi _\theta }}}}}$, $B_\theta ^* \in {\mathbb{R} ^{{n_{{\xi _\theta }}} \times {m_{{\xi _{}}}}}}$, $\Psi (.) \in {\mathbb{R} ^{{m_{{\xi _{}}}}}}$, and  ${\xi _\theta }(t) \in {\mathbb{R} ^{{n_{{\xi _\theta }}}}}$ denotes the nonlinear transformation of the state  $x$ through a vector-valued observable ${\xi _\theta }(t)$. The term $B_\theta ^*\Psi (.)$  captures how the observable  $\xi_{\theta} (t)$ is modified by the control policy $u(t)$.
Here, we assumed that ${\Psi}(.)$ is known. Note that this assumption is standard and realistic and it is about adding available knowledge (e.g., in the form of basis functions) to the design procedure (See \citep{korda2018linear,Brunton2016KoopmanIS}).
 Moreover, ${{\cal P}_\theta }(.):{\mathbb{R} ^{{n_{{\xi _\theta }}}}} \mapsto {\mathbb{R} ^{{n_{{\xi _\theta }}}}}$ is a bounded continuous approximation error term that depends on the goodness of the chosen library of observable functions. 
  {In the rest of the paper, for simplicity of notation, we have used  ${\xi _\theta }$ instead of ${\xi _\theta }(t)$, i.e., $t$ is dropped, when it is clear from the context.}




\begin{rem}\label{Remark:1}
Note that the dimension of the system \eqref{eq:16} scales with the number of observables, and it is desired to use a few dominant observables associated with persistent dynamics to reduce the size of the transformed system, make it amenable to control design methods. If the observables span contains $F(.)$, $G(.)$, and $x$, then the model \eqref{eq:16} may be well represented. {Additionally, it should be noted that the assumption of a bounded $\mathscr{P}_\theta(.)$ is a standard in the literature based on the universal approximator characteristics \citep{tao2003adaptive}.}
\end{rem}


The identifier for the unknown Koopman operators is defined as
\begin{align}
\frac{d}{{dt}}{\hat \xi _\theta } (t)= \hat A_\theta ^{}{\xi _\theta } (t) + \hat B_\theta ^{}\Psi (u(t)), \label{eq:17}
\end{align}
where ${\hat A_\theta (t)}$ and ${\hat B_\theta (t)}$  denote the identified matrices $A_\theta ^*$  and $B_\theta ^*$ at the time of $t$. Now, one can write the identification errors as
\begin{align}
\left\{ \begin{array}{l}
{e_{{\hat A_\theta }}}(t) = {\hat A_\theta }(t) - A_\theta ^*\\
{e_{{\hat B_\theta }}}(t) = {\hat B_\theta }(t) - B_\theta ^*\\
{e_{{\hat \xi _\theta }}}(t)= {\hat \xi _\theta }(t) - {\xi _\theta }(t)
\end{array} \right. \label{eq:18}
\end{align} 


The objective is now to design a unified batch-online update law for the system identifier \eqref{eq:17} to make the identification errors converge to zero, i.e.,  $\| {{e_{{\hat \xi _\theta }}}} \| \to 0$, $\| {{e_{{\hat A_\theta }}}} \| \to 0$, and  $\| {{e_{{\hat B_\theta }}}} \| \to 0$, in finite time for the case where there is no approximation error and the identification errors remain uniformly ultimately bounded (UUB) for the case where there is bounded approximation error. These objectives are formally stated in Problems 1 and 2 in the sequel. { We first consider the case with no approximation error term, i.e., ${{\cal P}_\theta }(.)=0$.} In the subsequent sections, then, the presence of the approximation error term will be analyzed. Moreover, the selection of an appropriate Koopman observables set that achieves the minimum approximation error and avoids corruption dynamics is performed through meta-learning later in the subsequent sections. 

{\bf Problem 1 (Batch-online Koopman finite-time identifier when there is no approximation error).} Let the vector-valued observable $\xi_\theta$ be fixed and result in no approximation error. Let exist a mini-batch of samples given by $\{\xi_{\theta}(t_{1}), \ldots, \xi_{\theta}(t_{p})\}$. Consider the  system \eqref{eq:16} along with the identifier \eqref{eq:17}. Develop a unified batch-online update law to make  $\| {{e_{{\hat \xi _\theta }}}} \| \to 0$, $\| {{e_{{\hat A_\theta }}}} \| \to 0$, and  $\| {{e_{{\hat B_\theta }}}} \| \to 0$, in finite time, under an easy-to-verify condition on mini-batch of collected data samples.

Letting the approximation error be zero, rewrite the system \eqref{eq:16} as
\begin{align}
\frac{d}{{dt}}{\xi _\theta } (t) = {\Sigma^*_{\theta }}^{\top}{{\cal Z}_\theta }({\xi _\theta },u(t)), \label{eq:19}
\end{align}	 
where $\Sigma {_{\theta }^{*\top}} = [A_\theta ^*,B_\theta ^*] \in {\mathbb{R} ^{{n_{{\xi _\theta }}} \times ({n_{{\xi _\theta }}} + {m_{{\xi _\theta }}})}}$ is the unknown matrix, and ${{\cal Z}_\theta }({\xi _\theta },u(t)) = {[{\xi _\theta ^{\top} (t)}, {\Psi _{}^{\top}(u(t))}]^{\top}} \in {\mathbb{R} ^{({n_{{\xi _\theta }}} + {m_{{\xi _{}}}})}}$.

{To eliminate the need to measure state derivatives as is required by existing system identification methods, the system model is formulated as a filtered regressor as follows.}

{
\begin{prop}\label{prop:2}
Consider the system \eqref{eq:16}, \eqref{eq:19}. This system can be expressed as the filtered form
\begin{align}
\left\{ \begin{array}{l}
{\xi _\theta } (t) = \Sigma _\theta ^*{h_\theta }({t }) + {a }l_\theta(t) + {e^{ - {a I_{n_{\xi _\theta }} }t}}{\xi _\theta }(0)\\
{{\dot h}_\theta }(t) =  - {a }{h_\theta }(t) + {{\cal Z}_\theta }({\xi _\theta },u(t)),\,\,\\
{{\dot l}_\theta }(t) =  - {a }{l_\theta }(t) + {\xi _\theta },\,\,\,\,\,\,\,\,\,\,\,\,\,\,\,\,\,\,\,\,\,\,\,\,\,\,\,\,\,\,
\end{array} \right. \label{eq:20}
\end{align}
$\forall {a } > 0$ with ${l_\theta }(t) = 0$ and ${h_\theta }(t) = 0$, where ${\xi _\theta }(0)$ denotes the initial state of  \eqref{eq:19} and  ${h_\theta }(t) \in {\mathbb{R} ^{({n_{{\xi _\theta }}} + {m_\xi })}}$  and  ${l_\theta }(t) \in {\mathbb{R} ^{{n_{{\xi _\theta }}}}}$ denote the filtered regressor form of ${{\cal Z}_\theta }({\xi _\theta },u)$  and  ${\xi _\theta }$, respectively. 
\end{prop}}
{
{\bf Proof.} The proof is similar to that of Lemma 1 in \citep{modares2013adaptive} and  it is provided here for the sake of completeness.
%
By adding and subtracting $a{\xi _\theta }$ from the right-hand side of \eqref{eq:19}, we get
\begin{align}
	\frac{d}{{dt}}{\xi _\theta }(t) = -a{\xi _\theta} (t)+ {\Sigma^*_{\theta }}^{\top}{{\cal Z}_\theta }({\xi _\theta },u(t))+a{\xi _\theta}(t). \label{eq:19P}
\end{align}	 
One can see that \eqref{eq:19P} has the following solution
\begin{align}
	 {\xi _\theta }(t)= {\Sigma^*_{\theta }}^{\top} \int_0^t e^{-aI_{n_{\xi _\theta }}(t-\tau)} {{\cal Z}_\theta }({\xi _\theta }(\tau), u(\tau)) d \tau 
	  +a \int_0^t e^{-aI_{n_{\xi _\theta }}(t-\tau)} {\xi _\theta }(\tau) d \tau+e^{-aI_{n_{\xi _\theta }} t} {\xi _\theta }(0).  \label{eq:19P1}
\end{align}
Define
\begin{align}
	{h_\theta }(t) &:=  \int_0^t e^{-aI_{n_{\xi _\theta }}(t-\tau)} {{\cal Z}_\theta }({\xi _\theta }(\tau), u(\tau)) d \tau,  \label{eq:19P2}\\
     {l_\theta }(t) &:=  \int_0^t e^{-aI_{n_{\xi _\theta }}(t-\tau)} {\xi _\theta }(\tau) d \tau. \label{eq:19P3}
\end{align}
Using \eqref{eq:19P1}-\eqref{eq:19P3}, we have
\begin{align}
	{\xi _\theta }(t) = \Sigma _\theta ^*{h_\theta }(t) + {a }l_\theta(t) + {e^{ - {a I_{n_{\xi _\theta }} }t}}{\xi _\theta }(0),
\end{align}
which is the first equation in  \eqref{eq:20}.
The second and third equations in \eqref{eq:20} can be obtained by taking the derivatives of \eqref{eq:19P2} and \eqref{eq:19P3}. This completes the proof.
\hfill $\square$}


Taking \eqref{eq:20} and dividing its both sides by the normalizing signal ${n_{{s_\theta }}} = 1 + {h_\theta }^{\top}(t){h_\theta }(t) + {l_\theta }^{\top}(t){l_\theta }(t)$, one has
\begin{align}
{{\bar \xi} _\theta }(t) = \Sigma {_{\theta }^{*\top}}{\bar h_\theta }(t) + {a }{\bar l_\theta }(t) + {e^{ -  {a I_{n_{\xi _\theta }} }t}}{{\bar \xi} _\theta }(0), \label{eq:21}
\end{align} 
where ${{\bar \xi} _\theta } = {{{\xi _\theta }} \mathord{\left/
 {\vphantom {{{\xi _\theta }} {{n_{{s_\theta }}}}}} \right.
 \kern-\nulldelimiterspace} {{n_{{s_\theta }}}}}$, {${{\bar \xi} _\theta }(0) = {{{\xi _\theta }}(0) \mathord{\left/
 {\vphantom {{{\xi _\theta }} {{n_{{s_\theta }}}}}} \right.
 \kern-\nulldelimiterspace} {{n_{{s_\theta }}}}}$}, ${\bar h_\theta }(t) = {{{h_\theta }(t)} \mathord{\left/
 {\vphantom {{{h_\theta }(t)} {{n_{{s_\theta }}}}}} \right.
 \kern-\nulldelimiterspace} {{n_{{s_\theta }}}}}$, and ${\bar l_\theta }(t) = {{{l_\theta }(t)} \mathord{\left/
 {\vphantom {{{l_\theta }(t)} {{n_{{s_\theta }}}}}} \right.
 \kern-\nulldelimiterspace} {{n_{{s_\theta }}}}}$ are the normalized forms of ${\xi _\theta }$, ${h_\theta }(t)$, and ${l_\theta }(t)$, respectively.

{Using \eqref{eq:21}, and Proposition \ref{prop:2}, the identifier's state \eqref{eq:17} becomes }
\begin{align}
{\hat {{\bar \xi}} _\theta } (t)= {\hat \Sigma _\theta }^{\top}(t){\bar h_\theta }(t) + {a }{\bar l_\theta }(t) + {e^{ -  {a I_{n_{\xi _\theta }} } t}}{{\bar \xi} _\theta }(0), \label{eq:22}
\end{align}
where ${\hat \Sigma _\theta }^{\top}(t): = [{{{\hat A}_\theta }(t)}, {{{\hat B}_\theta }(t)}
] \in {\mathbb{R} ^{{n_{{\xi _\theta }}} \times ({n_{{\xi _\theta }}} + {m_\xi })}}$.
The normalized version of identification error, i.e., ${\bar e_{{\xi _\theta }}} = {{{e_{{\xi _\theta }}}} \mathord{\left/
 {\vphantom {{{e_{{\xi _\theta }}}} {{n_{{s_\theta }}}}}} \right.
 \kern-\nulldelimiterspace} {{n_{{s_\theta }}}}}$, is defined as
\begin{align}
{\bar e_{{\xi _\theta }}}(t) = {\hat {\bar \xi} _\theta }(t) - {{\bar \xi} _\theta }(t) = {\tilde \Sigma _\theta }^{\top}{\bar h_\theta }(t), \label{eq:23}
\end{align} 
where  ${\tilde \Sigma _\theta }^{\top}(t) = {\hat \Sigma _\theta }^{\top} (t)- {\Sigma_\theta^{* }}^\top$ describes the Koopman operators' identification errors.
  One can see that ${\bar e_{{\xi _\theta }}}$ is measurable at each time since it is a function of the observables and states of the system are assumed to be measurable.
  Furthermore, ${\bar e_{{\xi _\theta }}}$ linearly relates to the state of the filtered regressor and identification errors of Koopman operators.
   Using this formulation, as we see later, one can reuse the recorded data in the update law without having to compute the derivatives of the system states.

Now, \eqref{eq:23} can be rewritten as
\begin{align}
{\bar e_{{\xi _\theta }}} (t)= ({\bar h_\theta }^{\top}(t) \otimes {I_{{n_{{\xi _\theta }}}}})^\top\,\tilde \Sigma _\theta ^{vec}(t), \label{eq:24}
\end{align}
where $({\bar h_\theta }^{\top}(t) \otimes {I_{{n_{{\xi _\theta }}}}}) \in {\mathbb{R} ^{{n_{{\xi _\theta }}}({n_{{\xi _\theta }}} + {m_\xi }) \times {n_{{\xi _\theta }}}}}$ and $\tilde \Sigma _\theta ^{vec} \in {\mathbb{R} ^{{n_{{\xi _\theta }}}({n_{{\xi _\theta }}} + {m_\xi })}}$. 

\begin{rem}\label{Remark:2}
Standard batch learning practice for parameter convergence of the Koopman identifier (i.e., for its generalization guarantees) \citep{han2020deep,netto2018robust} leverages statistical learning theory to provide probably approximately correct (PAC) sample complexity bounds on the learned model. PAC analysis, however, requires a huge number of i.i.d samples to guarantee generalization, which depends on the VC-dimension \citep{vapnik2013nature,blockeel2013machine}  of the search space. On the other hand, online learning algorithms guarantee parameter convergence (usually asymptotic guarantees in time) under restrictive and hard-to-verify persistence of excitation conditions. It is critical to design learning algorithms that bring the best of both worlds together and provide finite-time guarantees under easy-to-verify conditions (rather than i.i.d conditions) on samples that depend only on the dimension of the search space and not its VC-dimension.
\end{rem}

In this paper, we store a mini-batch of past samples in a history stack and retrieve them during online learning using the experience replay (also known as concurrent learning) technique \citep{chowdhary2010concurrent,modares2013adaptive}. This technique needs to collect past data in the history stack as
\begin{align}
{{\cal S}_\theta } = [{\bar h_\theta }({t_1}),\,...,\,{\bar h_\theta }({t_p})], \label{eq:25}
\end{align}
 $p$ is the number of collected data points that are stacked in the history stack, and ${t_1},\,...,\,{t_p}$ are their associated recorded times.
The error of identification for the $j$-th collected data point is calculated as follows
\begin{align}
{{\bar e}_{{\xi _\theta }}}(t,\,{t_j}) =& {{\hat {{\bar \xi}} }_\theta }(t,\,{t_j}) - {{{\bar \xi} }_\theta }({t_j}) \nonumber \\
 =& {{\tilde \Sigma }_\theta }^{\top}(t)\,{{\bar h}_\theta }({t_j}),\, \nonumber \\
 =& ({{\bar h}_\theta }^{\top}({t_j} ) \otimes {I_{{n_{{\xi _\theta }}}}})\,\tilde \Sigma _\theta ^{vec}(t),
 \label{eq:26}
\end{align} 
{
for $j = 1,\,...,\,p$, where ${{\bar \xi} _\theta }({t_j})$ denotes the normalized form of the state at  ${t_j}$, ${\hat {\bar \xi} _\theta }(t,\,{t_j})$ denotes identifier's state at ${t_j}$ defined as
\begin{align}
{\hat {{\bar \xi}} _\theta }(t,{t_j}) := {\hat \Sigma _\theta }^{\top}(t){\bar h_\theta }({t_j}) + {a }{\bar l_\theta }({{t_j}}) + {e^{ -  {a I_{n_{\xi _\theta }} } t_j}}{{\bar \xi} _\theta }(0),
	\label{eq:26nneeww}
\end{align}
 and  ${\bar e_{{\xi _\theta }}}(t,\,{t_j})$ is the error of identification at ${t_j}$.}
Furthermore, ${\tilde \Sigma _\theta }^{\top}(t)$ is the identification error of Koopman operators at the present time. 


\begin{cond}\label{cond:3}
The stacked data ${{\cal S}_\theta }$  at least consists of numbers of linearly independent elements equal to the dimension of the basis function ${h_\theta }(t)$. That is,  $\sum\limits_{j = 1}^p {{{\bar h}_\theta }({t_j})} \,{\bar h^{\top}}({t_j})\, \ge d_\theta{I_{{n_{{\xi _\theta }}} + {m_\xi }}}$ for some  $d_\theta \in \mathbb{R}^{+}$. 
\end{cond}


\section{Main result}

The hierarchical learning architecture given in Fig. 1 is adopted to deal with the problem at hand, which  consists of:\\
- A base learner (finite-time identifier), which is used to find the Koopman operator ${\hat A_\theta }(t)$  and ${\hat B_\theta }(t)$  corresponding to a library of observable function $\theta $.\\
- A meta-learner to learn the best library of observables $\theta $ that achieves a minimum approximation error based on the available collected data set.


\subsection{Base Learner: Batch-online Finite-time Koopman Identifier with no Approximation Error }
It is well known that solution trajectories of the locally Lipschitz continuous systems converge no faster than exponentially to equilibrium points, i.e., such systems at most can only have asymptotic convergence rates. 
Non-smooth or non-Lipschitz continuous systems, however, are able to have a finite-time convergence property.
 This emphasizes that enjoying the finite-time convergence property in continuous-time systems is feasible only by introducing update law with discontinuous or non-Lipschitz dynamics based on differential inclusions instead of ODEs.
  The following theorem provides a unified novel batch-online learning using a novel discontinuous flow of gradients to guarantee finite-time convergence for the Koopman identifier when there is no approximation error and the observables are fixed.

\begin{thm}\label{theorem:1}
Consider the systems \eqref{eq:24} and \eqref{eq:26}. Let Condition \ref{cond:3} and Assumption \ref{Assm:1} hold. Then, the unknown parameter vector  $\hat \Sigma _\theta ^{vec}(t)$ converges to its true ${\Sigma_\theta^{vec}}^*$  in finite time by using any maximal Filippov solution to the following discontinuous differential inclusion update law
{
\begin{align}
%
%
\dot {\hat \Sigma} _\theta ^{vec}(t) = \dot {\tilde \Sigma} _\theta ^{vec}(t) \in {{K_\theta } }=
{\bf K}[{ {\cal I}_\theta }]:{\mathbb{R} ^{{n_{{\xi _\theta }}}({n_{{\xi _\theta }}} + {m_\xi })}} \mathbin{\lower.3ex\hbox{$\buildrel\textstyle\rightarrow\over
		{\smash{\rightarrow}\vphantom{_{\vbox to.5ex{\vss}}}}$}} {\mathbb{R} ^{{n_{{\xi _\theta }}}({n_{{\xi _\theta }}} + {m_\xi })}},
 \label{eq:28}
\end{align}}
with {
\begin{align}
{\cal I_\theta } =  - {\alpha _\theta }\frac{\| {{\bf{H}}_\theta(t)} \|}{{{{( {{\bf{H}}_\theta(t)} )}^{\top}}}} 
 \frac{{{{\left[ {{{\bf{A}}_\theta }} \right]}^r} {{\bf{H}}_\theta(t)} }}{{{{\left[ {{{\bf{A}}_\theta }} \right]}^{r + 1}} {{\bf{H}}_\theta(t)} }},
 \label{eq:29}
\end{align}
where
\begin{align}
	{{\bf{H}}_\theta(t)} := {{{\cal H}_\theta }(t){{\bar e}_{{\xi _\theta }}}(t) + \sum\limits_{j = 1}^{{p_\theta }} {{{\cal H}_\theta }({t_j}){{\bar e}_{{\xi _\theta }}}(t,{\mkern 1mu} {t_j})} {\mkern 1mu} },
	\label{eq:31TT}
\end{align}}
and
\begin{align}
&{{\cal H}_\theta }(t) := {{\bar h}_\theta }(t) \otimes {I_{{n_{{\xi _\theta }}}}}, \nonumber \\
&{{\cal H}_\theta }({t_j}) := {{\bar h}_\theta }({t_j}) \otimes {I_{{n_{{\xi _\theta }}}}} , \label{eq:31}
\end{align}
\begin{align}
    {{\bf{A}}_\theta } := {{\cal H}_\theta }(t){{\cal H}_\theta }^{\top}(t) + \sum\limits_{j = 1}^{{p_\theta }} {{{\cal H}_\theta }({t_j}){{\cal H}_\theta }^{\top}({t_j})} , \label{eq:32a}
\end{align}
where {$r \in \mathbb{R} $}, ${\alpha _\theta } > 0$ and ${K_\theta } = {\bf K}[{ {\cal I}_\theta }]:{\mathbb{R} ^{{n_{{\xi _\theta }}}({n_{{\xi _\theta }}} + {m_\xi })}} \mathbin{\lower.3ex\hbox{$\buildrel\textstyle\rightarrow\over
{\smash{\rightarrow}\vphantom{_{\vbox to.5ex{\vss}}}}$}} {\mathbb{R} ^{{n_{{\xi _\theta }}}({n_{{\xi _\theta }}} + {m_\xi })}}$  is a set-valued map. Furthermore, its convergence time is given by the exact settling time
{
\begin{align}
{t_\theta }^ \star  = \frac{1}{{{\alpha _\theta }}}{\| {{\bf{H}}_\theta(0)} \|}. \label{eq:32}
\end{align}}
\end{thm}

\begin{figure}[!t]
\centering{\includegraphics [width=5.5in] {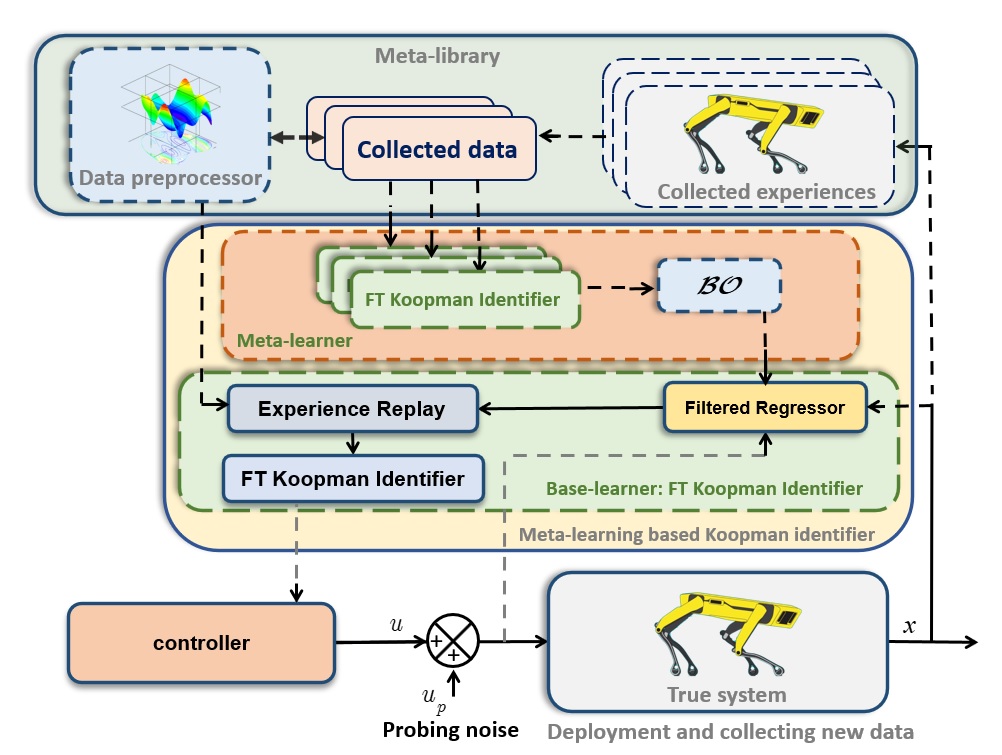}}
\caption{Proposed hierarchical learning architecture using Bayesian optimization (BO) for automatic lifted-state tuning. The lifted-state configuration ${\theta _k}$  is evaluated in the meta-learner in terms of cost functional ${J^R}(.)$. Based on all previous experiments, BO suggests the next lifted-state configuration ${\theta _{k + 1}}$  aiming at finding the global optimum with only a few iterations.} \label{fig:1}
\end{figure}

\noindent {\bf Proof.} {  It follows from \eqref{eq:24}, \eqref{eq:26}, \eqref{eq:31TT}, and \eqref{eq:31} that
\begin{align}
		{{\bf{H}}_\theta(t)} = {{\bf{A}}_\theta }\tilde \Sigma _\theta ^{vec}(t).
		\label{eq:32aaa}
\end{align} 
Using \eqref{eq:32aaa}, \eqref{eq:29} can be rewritten as}
{\begin{align}
{{\cal I}_\theta } =  - {\alpha _\theta }\frac{{\|{{\bf{A}}_\theta }\tilde \Sigma _\theta ^{vec}\|}}{{\tilde \Sigma {{_\theta ^{vec}}^{\top}} {{{\bf{A}}_\theta }} }}\,\frac{{{{\left[ {{{\bf{A}}_\theta }} \right]}^r} {{{\bf{A}}_\theta }} \tilde \Sigma _\theta ^{vec}}}{{{{\left[ {{{\bf{A}}_\theta }} \right]}^{r + 1}} {{{\bf{A}}_\theta }} \tilde \Sigma _\theta ^{vec}}}. \label{eq:33}
\end{align}}
We begin now by proving that Condition \ref{cond:1} is satisfied. That is, ${K_\theta } = {\bf K}[{{\cal I}_\theta }]:{\mathbb{R} ^{{n_{{\xi _\theta }}}({n_{{\xi _\theta }}} + {m_\xi })}} \mathbin{\lower.3ex\hbox{$\buildrel\textstyle\rightarrow\over
{\smash{\rightarrow}\vphantom{_{\vbox to.5ex{\vss}}}}$}} {\mathbb{R} ^{{n_{{\xi _\theta }}}({n_{{\xi _\theta }}} + {m_\xi })}}$ is upper semi-continuous, and has compact, nonempty, and convex values. To this aim, using \eqref{eq:33} and some manipulation, give
\begin{align}
\left\| {{{\cal I}_\theta }} \right\|\,\,\,\, &\le {\alpha _\theta }\|\frac{1}{2}{{\bf{A}}_\theta }\tilde \Sigma _\theta ^{vec}\,\| \frac{{{\lambda _{\max }}({{[{{\bf{A}}_\theta }]}^r})}}{{{\lambda _{\min }}({{[{{\bf{A}}_\theta }]}^{r + 1}})}}\frac{{\left\| {\frac{1}{2}{{\bf{A}}_\theta }\tilde \Sigma _\theta ^{vec}} \right\|}}{{{{\left\| {\frac{1}{2}{{\bf{A}}_\theta }\tilde \Sigma _\theta ^{vec}} \right\|}^2}}} \nonumber \\
 &\le {\alpha _\theta }\frac{{{\lambda _{\max }}({{[{{\bf{A}}_\theta }]}^r})}}{{{\lambda _{\min }}({{[{{\bf{A}}_\theta }]}^{r + 1}})}}.
 \label{eq:34}
\end{align}
Based on Condition \ref{cond:3},
\begin{align}
\sum\limits_{j = 1}^{{p_\theta }} {{{\cal H}_\theta }({t_j}){{\cal H}_\theta }^{\top}({t_j}) = }  \sum\limits_{j = 1}^{{p_\theta }} ( {{\bar h}_\theta }({t_j})  {{\bar h}_\theta }^{\top}({t_j}) ) \otimes {I_{{n_{{\xi _\theta }}}}} \ge  d_\theta {I_{{n_{{\xi _\theta }}}({n_{{\xi _\theta }}} + {m_\xi })}}. \label{eq:33a}
\end{align}
Hence, $\Scale[1]{{{\cal H}_\theta }(t){{\cal H}_\theta }^{\top}(t)  \ge {0_{{n_{{\xi _\theta }}}({n_{{\xi _\theta }}} + {m_\xi })}}}$ implies that $\Scale[0.94]{{\lambda _{\min }}(({{\cal H}_\theta }(t){\cal H}_\theta ^ \top (t) + \allowbreak \sum\limits_{j = 1}^{{p_\theta }} {{{\cal H}_\theta }} \left( {{t_j}} \right){\cal H}_\theta ^ \top \left( {{t_j}} \right){)^{r + 1}})} \ge {d_\theta }$ everywhere near  $\hat \Sigma _\theta ^{vec} = {\Sigma_\theta ^{vec}}^*$ for some  ${d_\theta } > 0$.
{
 As a result, \eqref{eq:34} implies that ${{\cal I}_\theta }$
   is  defined a.e. and is Lebesgue measurable in a non-empty open region ${\cal R} \subset {\mathbb{R} ^{{n_{{\xi _\theta }}}({n_{{\xi _\theta }}} + {m_\xi })}}$, and for every point  $\tilde \Sigma _\theta ^{vec} \in {\mathbb{R} ^{{n_{{\xi _\theta }}}({n_{{\xi _\theta }}} + {m_\xi })}}$, ${{\cal I}_\theta }$ is bounded a.e. in some bounded neighborhood of  $\tilde \Sigma _\theta ^{vec}$.
 }
 Using this observation and { Theorem 5 of \citep[Chapter~2]{filippov1988differential}}, one can conclude that Condition \ref{cond:1} is satisfied, i.e., the Filippov set-valued map ${K_\theta } = {\bf K}[{{\cal I}_\theta }]$  
 is upper semi-continuous, and has compact, nonempty, and convex values.

Now, take the following continuously differentiable Lyapunov function defined over ${\mathbb{R} ^{{n_{{\xi _\theta }}}({n_{{\xi _\theta }}} + {m_\xi })}}$, which w.r.t. $\tilde \Sigma _\theta ^{vec}$ is positive definite.
{
\begin{align}
V(\tilde \Sigma _\theta ^{vec}) &= {\tilde \Sigma_\theta ^{vec  \top}} {\left( {{{\bf{A}}_\theta }} \right)^2}\tilde \Sigma_\theta ^{vec} \nonumber \\
&= 
\left\|\mathbf{A}_\theta \tilde{\Sigma}_\theta^{v e c}\right\|^2.
 \label{eq:35}
\end{align}}
Despite the update law \eqref{eq:28}-\eqref{eq:29} is continuous close to  ${\Sigma_\theta^{vec}}^*$, it is not continuous at ${\Sigma_\theta^{vec}}^*$,  and therefore, undefined at $\hat \Sigma _\theta ^{vec} = {\Sigma_\theta^{vec}}^*$  itself. To this end, given $\tilde \Sigma _\theta ^{vec}(t) \in {\mathbb{R} ^{{n_{{\xi _\theta }}}({n_{{\xi _\theta }}} + {m_\xi })}}\backslash \{ {\Sigma_\theta^{vec}}^*\} ,$  one has
\begin{align}
\begin{array}{l}
\sup \dot V(\tilde \Sigma _\theta ^{vec})= \sup \{ {a_\theta } \in \mathbb{R} :\exists v \in {\bf K}[{{\cal I}_\theta }](\tilde \Sigma _\theta ^{vec})\,s.t.\,\,{a_\theta } = p \cdot v,\,\forall p \in \partial V(\tilde \Sigma _\theta ^{vec})\} \\
\quad \quad \quad \quad \quad = \sup \left\{ {2\,{{({{\bf{A}}_\theta })}^2}\tilde \Sigma _\theta ^{vec} \cdot \left( {{{\cal I}_\theta }(\tilde \Sigma _\theta ^{vec}(t))} \right)} \right\}\\
\quad \quad \quad \quad \quad =  - 2{\alpha _\theta }\|{{\bf{A}}_\theta }\tilde \Sigma _\theta ^{vec}\|\\
\quad \quad \quad \quad \quad =  - (2{\alpha _\theta })(V(\tilde \Sigma _\theta ^{vec}))^{\frac{1}{2}}.
\end{array} \label{eq:36}
\end{align}
Note that \eqref{eq:35} satisfies Condition \ref{cond:2} since it is a continuously differentiable function. Furthermore, for $\hat \Sigma _\theta ^{vec} = {\Sigma_\theta^{vec}}^*$, i.e., $\tilde \Sigma _\theta ^{vec}(t) = 0$,  $\dot V\left( 0 \right) = 0$ since
$${\nabla _{\tilde \Sigma _\theta ^{vec}}}V\left( 0 \right) = 2({({{\cal H}_\theta }(t){{\cal H}_\theta }^{\top}(t) + \sum\limits_{j = 1}^{{p_\theta }} {{{\cal H}_\theta }({t_j}){{\cal H}_\theta }^{\top}({t_j}))} )^2}\tilde \Sigma _\theta ^{vec} = 0.$$
 Moreover, using \eqref{eq:28}-\eqref{eq:31}, one has
{
\begin{align}
V(\tilde \Sigma _\theta ^{vec}) = {( {{{\bf{H}}_\theta(t)}})^{\top}} 
{{{\bf{H}}_\theta(t)}}, \label{eq:37}
\end{align}
which implies that 
\begin{align}
V( {\tilde \Sigma _\theta ^{vec}(0)} ) = {( {{{\bf{H}}_\theta(0)}} )^{\top}} 
 {{{\bf{H}}_\theta(0)}}. \label{eq:37a}
\end{align}}
Now, invoking Proposition \ref{prop:1} and observing that $\sup \dot V(\tilde \Sigma _\theta ^{vec}) = - (2\alpha )V{(\tilde \Sigma _\theta ^{vec})^{\frac{1}{2}}}$  and $V\left( {\tilde \Sigma _\theta ^{vec}(0)} \right)$  is computable, maximal Filippov solution to the discontinuous differential inclusion update law \eqref{eq:28}-\eqref{eq:29} converges to  ${\Sigma_\theta^{vec}}^*$ in finite time by the exact prescribed settling time \eqref{eq:32}. Using \eqref{eq:24}, 
the convergence of ${\tilde \Sigma _\theta ^{vec}}$ within a finite time implies that ${e_{{\xi _\theta }}}$ also goes to the origin within a finite time.
 This completes the proof. \hfill $\square$

{
\begin{rem}\label{Remark:30}
From \eqref{eq:34}, one can see that a history stack's richness influences the convergence time of the identification error. To check whether finite-time convergence is possible, condition 3 provides an easy-to-verify rank condition. One can add a probing noise into the control signal to ensure that Condition 3 is satisfied. 
\end{rem}}

{
\begin{rem}\label{Remark:3}
It is also worth noting that even though the number of stored samples in the history stack is fixed after Condition 3 is satisfied, replacing old data with new rich data in the history stack can avoid numerical challenges in algorithm iterations in the case of poorly conditioned $\mathbf{A}_\theta$. To achieve this, new samples are periodically added, and old ones are removed if ${{\lambda _{\min }}({{[{{\bf{A}}_\theta }]}^{r+1}})}$ would be increased. This method, however, requires new data samples stream in. Therefore, in meanwhile,
to keep the update law \eqref{eq:28}-\eqref{eq:29} practical in terms of being suitable for numerical solvers, one can also incorporate a small constant regularization term  $\eta  > 0$ (which is a small constant) in \eqref{eq:29}, as
\begin{align}
{{\cal I}_\theta } =  - {\alpha _\theta }\frac{{\|{{{\bf{H}}_\theta(t)}}\|}}{{\eta  + {{{\bf{H}}_\theta(t)}}}} \frac{{{{\left[ {{{\bf{A}}_\theta }} \right]}^r}{{{\bf{H}}_\theta(t)}}}}{{{{\left[ {{{\bf{A}}_\theta }} \right]}^{r + 1}}{{{\bf{H}}_\theta(t)}}}},
\label{eq:38}
\end{align}
and then remove it when $\mathbf{A}_\theta$ is not poorly conditioned anymore.
It is worth noting that the addition of this term transforms finite-time convergence into  practical finite-time convergence. That is, one can select sufficiently small $ \eta  $  to go into any given arbitrary neighborhood of the origin.
Note that the concept of practical finite-time convergence has recently received significant attention \citep{liu2021semi,chen2021practical,chowdhary2011singular}  since it specifies a convergence bound and finite convergence time that can be effectively employed in control, identification, and monitoring to improve performance and reduce conservatism in control design.
\end{rem}}

\begin{rem}\label{Remark:4}
To assure $\|e_{{A_\theta}}\| \to 0$ and $\|e_{{B_\theta}}\| \to 0$, the presented update law leverages modified gradient descent rule to select ${\hat \Sigma}_{vec} ^{\theta}$ to minimize the following cost function for not only  the current time, but also the past samples collected in the memory.
\begin{align}
J(t)= {\bar e_{{\xi _\theta }}(t)}^{\top}{\bar e_{{\xi _\theta }}(t)}+\sum_{j=1}^{p_\theta} {\bar e_{{\xi _\theta }}}(j)^{\top}{\bar e_{{\xi _\theta}}(j)}, \label{eq:RE4}
\end{align}
where its gradient and the Hessian matrix can be calculated as 
\begin{align}
{\nabla _{\tilde \Sigma _\theta ^{vec}}}J = \frac{1}{2}\left( {{{\bf{A}}_\theta } + {\bf{A}}_\theta ^{\top}} \right)\tilde \Sigma _\theta ^{vec} = {{\bf{A}}_\theta } \tilde \Sigma _\theta ^{vec}, \label{eq:RE4a}
\end{align}
\begin{align}
\nabla _{\tilde \Sigma _\theta ^{vec}}^2J = \frac{1}{2}{{\bf{A}}_\theta } + \frac{1}{2}{\bf{A}}_\theta ^{\top} = {{\bf{A}}_\theta }. \label{eq:RE4b}
\end{align}
\end{rem}
Since ${\bf{A}}_\theta $ in \eqref{eq:RE4a} depends on both current (online) and past (batch) samples, the discontinuous gradient update law minimizes the identification error for both current and past samples. Before the rank condition is satisfied, only estimation error is guaranteed to converge to zero (no generalization guarantee), and as the samples are collected to satisfy Condition 3, the data samples provide a good representation of the dynamic system \eqref{eq:16} across its entire operating regimes, if an appropriate set of observables is chosen (which will be performed in a meta-layer in this paper). 


We now analyze the efficiency of the update law \eqref{eq:28}-\eqref{eq:29} by using the notion of regret. 
To this aim, let the normed error ${\cal N}(.)$ and continuous regret be defined as
\begin{align}
\begin{array}{l}
{\cal N}(\tilde \Sigma _\theta ^{vec}(t)) = \| {{{{\bf{A}}_\theta }}(t)\tilde \Sigma _\theta ^{vec}(t)} \|
\end{array} \label{eq:52}
\end{align} 
where 
\begin{align}
\begin{array}{l}
Regret: = \int_0^{{t_\theta }^*} {{\cal N}(\tilde \Sigma _\theta ^{vec}(\tau ))} d\tau  - {\min _{\hat \Sigma _\theta ^{vec} \in {\Theta _\theta }}}\int_0^{{t_\theta }^*} {\cal N} (\tilde \Sigma _\theta ^{vec}(\tau ))d\tau \\
\,\,\,\,\,\,\,\,\,\,\,\, \quad \quad  = \int_0^{{t_\theta }^*} {{\cal N}(\tilde \Sigma _\theta ^{vec}(\tau ))} d\tau 
\end{array} \label{eq:54}
\end{align}
Note that ${\min _{\hat \Sigma _\theta ^{vec} \in {\Theta _\theta }}}\int_0^{{t_\theta }^*} {\cal N} (\tilde \Sigma _\theta ^{vec}(\tau ))d\tau  = 0$, since ${\min _{\hat \Sigma _\theta ^{vec} \in {\Theta _\theta }}}{\cal N}(\tilde \Sigma _\theta ^{vec}(t)) = 0$ and regret given in \eqref{eq:54} is a non-decreasing function of the time span ${t_\theta }^*$ since it contains a sum of ${\cal N}(\tilde \Sigma _\theta ^{vec}(t))$, which are non-negative costs. 

\begin{thm}\label{theorem:3}
Consider the regret given in \eqref{eq:54}. Regret grows sub-linearly with time and the convergence of identification errors to zero for the update rule \eqref{eq:28}-\eqref{eq:29}  is upper bounded by a constant regret $\frac{1}{{2{\alpha _\theta }}}V(0)$ where $V(.)$ is given in \eqref{eq:35}.
\end{thm}
{\bf Proof.} 
Regret is analyzed based on the Lyapunov stability condition provided in Theorem 1 by $\dot V(t) \le  - (2{\alpha _\theta })\| {{{{\bf{A}}_\theta }}(t)\tilde \Sigma _\theta ^{vec}(t)} \| \le 0$. Based on \eqref{eq:35}, one has $V(t) \ge 0$. Now, integrating $\dot V(t)$  from $0$  to  ${t_\theta }^*$, one has
\begin{align}
\int_0^{{t_\theta }^*} {\| {{{{\bf{A}}_\theta }}(\tau )\tilde \Sigma _\theta ^{vec}(\tau )} \|} d\tau  \le - \frac{1}{{2{\alpha _\theta }}}\int_0^{{t_\theta }^*} {\dot V} (\tau )d\tau  = \frac{1}{{2{\alpha _\theta }}}(V(0) - V({t_\theta^*})) = \frac{1}{{2{\alpha _\theta }}}V(0). \label{eq:55}
\end{align}               
Given that $\dot V(t) \le 0$, it can be seen that $V(0) - V({t^*}) \le V(0) = {\cal O}(1)$. This implies that the regret bound does not grow as a function of time (i.e., regret grows sub-linearly with time) since the convergence of identification errors to zero for the update rule \eqref{eq:28}-\eqref{eq:29}, which is upper bounded by a constant regret $\frac{1}{{2{\alpha _\theta }}}V(0)$. This completes the proof.  \hfill $\square$

\subsection{Batch-online Koopman finite-time identifier with approximation error}
 
Now, we investigate the sensitivity of the proposed update law \eqref{eq:28}-\eqref{eq:29} to the approximation error term ${\cal P}_\theta(.)$ by studying the behavior of solutions of the system identifier \eqref{eq:17} in a neighborhood of the finite-time solution of the nominal Koopman identifier. Let's assume that there exists the approximation error term ${\cal P}_\theta(.)$. 

{\bf Problem 2 (Batch-online Koopman finite-time identifier with approximation error).} Let the vector-valued observable $\xi (\theta )$ be fixed and results in some approximation error. Let the exist a mini-batch of samples given by $\{\xi_{\theta}(t_{1}), \ldots, (\xi_{\theta}(t_{p})\}$. 
Considering \eqref{eq:16} and the system identifier \eqref{eq:17}, develop a unified batch-online update law to ensure that ${e_{{\xi _\theta }}} $, ${e_{{A_\theta }}} $, and  ${e_{{B_\theta }}} $, are uniformly ultimately bounded (UUB),
 under an easy-to-verify condition on mini-batch of samples.

\begin{assm}\label{Assm:4}
Continuous approximation error term ${{\cal P}_\theta }(t,{\xi _\theta }(t))$ is bounded, i.e., $\parallel {{\cal P}_\theta }(t,{\xi _\theta }(t))\parallel \, \le {\cal P}_\theta ^B$.
\end{assm}

Rewrite the system \eqref{eq:19} as
\begin{align}
\frac{d}{{dt}}{\xi _\theta }(t) = {\Sigma _\theta^*}^{\top}{{\cal Z}_\theta }({\xi _\theta },u) + {{\cal P}_\theta }(t,{\xi _\theta }(t)). \label{eq:40}
\end{align}

{
	\begin{prop}\label{prop:3}
		Consider the system \eqref{eq:40}. This system can be expressed as the filtered form
		\begin{align}
			\left\{ \begin{array}{l}
				{\xi _\theta } (t)= {\Sigma _\theta^* }^{\top}{h_\theta }(t) + {a }{l_\theta }(t) + {e^{ -  {a I_{n_{\xi _\theta }} } t}}{\xi _\theta }(0) + {{\cal P}_\theta }(t)\\
				{{\dot h}_\theta }(t) =  - {a_\theta }{h_\theta }(t) + {{\cal Z}_\theta }({\xi _\theta },u),\,\,{h_\theta }({\xi _\theta }(0)) = 0\\
				{{\dot l}_\theta }(t) =  - {a }{l_\theta }(t) + {\xi _\theta },\,\,\,\,\,\,\,\,\,\,\,\,\,\,\,{l_\theta }({\xi _\theta }(0)) = 0
			\end{array} \right. \label{eq:41}
		\end{align}
		$\forall {a } > 0$ with ${l_\theta }(0) = 0$ and ${h_\theta }(0) = 0$, where ${{\cal P}_\theta } = \int\limits_0^{T} {{e^{ - { {a I_{n_{\xi _\theta }} } }(t - \tau )}}{{\cal P}_\theta }(\tau ,{\xi _\theta }(\tau ))d\tau } $ and ${\xi _\theta }(0)$ denotes the initial state of  \eqref{eq:40} and  ${h_\theta }(t) \in {\mathbb{R} ^{({n_{{\xi _\theta }}} + {m_\xi })}}$  and  ${l_\theta }(t) \in {\mathbb{R} ^{{n_{{\xi _\theta }}}}}$ denote the filtered regressor form of ${{\cal Z}_\theta }({\xi _\theta },u)$  and  ${\xi _\theta }$, respectively. 
\end{prop}}
{
	{\bf Proof.} The proof follows from similar development as Proposition 2.}


Using Proposition \ref{prop:3}, \eqref{eq:23} is reformulated as
 \begin{align}
\left\{ \begin{array}{l}
{{\bar e}_{{\xi _\theta }}} (t)= {{\tilde \Sigma }_\theta }^{\top}{{\bar h}_\theta }(t) - {{\bar {\cal P}}_\theta }(t)\\
{\xi _\theta }(t) = {\Sigma _\theta^* }^{\top}{h_\theta }(t) + {a }{l_\theta }(t) + {e^{ - { {a I_{n_{\xi _\theta }} } }t}}{\xi _\theta }(0) + {{\cal P}_\theta }(t)\\
{{\hat {\bar \xi} }_\theta } (t)= {{\hat \Sigma }_\theta }^{\top}{{\bar h}_\theta }(t) + {a }{{\bar l}_\theta }(t) + {e^{ - { {a I_{n_{\xi _\theta }} } }t}}{{\bar \xi }_\theta }(0)\\
{{\dot h}_\theta }(t) =  - {a_\theta }{h_\theta }(t) + {{\cal Z}_\theta }({\xi _\theta },u),\,\,\,\,{h_\theta }({\xi _\theta }(0)) = 0\\
{{\dot l}_\theta }(t) =  - {a }{l_\theta }(t) + {\xi _\theta },\,\,\,\,\,\,\,\,\,\,\,\,\,\,\,\,\,\,\,{l_\theta }({\xi _\theta }(0)) = 0
\end{array} \right. \label{eq:42}
\end{align}
where ${\bar {\cal P}_\theta }= {{{{\cal P}_\theta }} \mathord{\left/
 {\vphantom {{{{\cal P}_\theta }} {{n_{{s_\theta }}}}}} \right.
 \kern-\nulldelimiterspace} {{n_{{s_\theta }}}}}$.

Now, \eqref{eq:24} and \eqref{eq:26} are rewritten as
 \begin{align}
&{\bar e_{{\xi _\theta }}}(t) = ({\bar h_\theta }^{\top}(t) \otimes {I_{{n_{{\xi _\theta }}}}})\,\tilde \Sigma _\theta ^{vec} (t)- {\bar {\cal P}_\theta } (t)\label{eq:43} \\
&{\bar e_{{\xi _\theta }}}(t,\,{t_j}) = ({\bar h_\theta }^{\top}({t_j}) \otimes {I_{{n_{{\xi _\theta }}}}})\,\tilde \Sigma _\theta ^{vec}(t) - {\bar {\cal P}_\theta }(t),\,j = 1,\,...,\,{p_\theta }. \label{eq:44}
\end{align}

The following theorem concerns the behavior of identifier error (41) along with the update law \eqref{eq:28}-\eqref{eq:29} under finite-dimensional Koopman with bounded approximation error.

\begin{thm}\label{theorem:2}
Consider \eqref{eq:16} and \eqref{eq:41}, and let Assumptions \ref{Assm:1} and \ref{Assm:4} and Condition \ref{cond:3} hold.
In the case that a bounded Koopman approximation error exists, the update law \eqref{eq:28}-\eqref{eq:29} ensures that all the identification errors are UUB.
\end{thm}
{\bf Proof.} Using \eqref{eq:29}, \eqref{eq:42}, \eqref{eq:43}, and some  manipulations, one has
\begin{align}
\dot {\hat \Sigma} _\theta ^{vec}(t) = \dot {\tilde \Sigma} _\theta ^{vec}(t) \in {{K_\theta } }=
{\bf K}[{ {\cal I}_\theta }]:{\mathbb{R} ^{{n_{{\xi _\theta }}}({n_{{\xi _\theta }}} + {m_\xi })}} \mathbin{\lower.3ex\hbox{$\buildrel\textstyle\rightarrow\over
		{\smash{\rightarrow}\vphantom{_{\vbox to.5ex{\vss}}}}$}} {\mathbb{R} ^{{n_{{\xi _\theta }}}({n_{{\xi _\theta }}} + {m_\xi })}},
 \label{eq:45}
\end{align}
where
{
\begin{align}
\begin{array}{l}
{{\cal I}_\theta }(t,\tilde \Sigma _\theta ^{vec},{{\bar {\cal P}}_\theta }) =  - {\alpha _\theta }\|{{\bf{A}}_\theta }\tilde \Sigma _\theta ^{vec} - {{\rm{B}}_\theta }{{\bar {\cal P}}_\theta }\|  \Gamma _\theta ,
\end{array} \label{eq:46}
\end{align}
with
\begin{align}
\Gamma _\theta 	= \frac{{{{\left[ {{{\bf{A}}_\theta }} \right]}^r}\left[ { {{{\bf{A}}_\theta }} \tilde \Sigma _\theta ^{vec} - {{\rm{B}}_\theta }{{\bar {\cal P}}_\theta }} \right]}}{{{{\left[ { {{{\bf{A}}_\theta }} \tilde \Sigma _\theta ^{vec} - {{\rm{B}}_\theta }{{\bar {\cal P}}_\theta }} \right]}^{\top}}{{\left[ {{{\bf{A}}_\theta }} \right]}^{r + 1}}\left[ { {{{\bf{A}}_\theta }} \tilde \Sigma _\theta ^{vec} - {\rm{B}}_\theta {{\bar {\cal P}}_\theta }} \right]}},
\end{align}
\begin{align}
{{\rm{B}}_\theta } := {{\cal H}_\theta }(t) + \sum\limits_{j = 1}^{{p_\theta }} {{{\cal H}_\theta }({t_j})}. \label{eq:47}
\end{align}
Note that analysis to show the validity of Conditions 1 and 2 is similar to that in the proof of Theorem 1. Using \eqref{eq:45}-\eqref{eq:47} and Definition \ref{defn:21}, the derivative of continuously differentiable Lyapunov function \eqref{eq:35} becomes
\begin{align}
\sup \dot V(\tilde \Sigma _\theta ^{vec}) &= \sup \{ {a_\theta } \in \mathbb{R} :\exists v \in {\bf K}[{{\cal I}_\theta }](\tilde \Sigma _\theta ^{vec})\,s.t.\,\,{a_\theta } = p \cdot v,\,\forall p \in \partial V(\tilde \Sigma _\theta ^{vec})\} \nonumber \\
&= \sup \{ 2{( {{{\bf{A}}_\theta }} )^2}  {{\tilde \Sigma} _\theta ^{vec}} \cdot ( {{\cal I}_\theta }(t,{\tilde \Sigma} _\theta ^{vec},{{\bar {\cal P}}_\theta }) ) \} \nonumber \\
&=  - 2{\alpha _\theta }{\left( {{{\bf{A}}_\theta }} \right)^2}\tilde \Sigma _\theta ^{vec} \left\| { {{{\bf{A}}_\theta }} \tilde \Sigma _\theta ^{vec} - {{\rm{B}}_\theta }{{\bar {\cal P}}_\theta }} \right\|  
\Gamma _\theta .
\end{align}
Adding and subtracting $- 2{\alpha _\theta } {{{\bf{A}}_\theta }} {{{\rm{B}}_\theta }{{\bar {\cal P}}_\theta }} \left\| { {{{\bf{A}}_\theta }} \tilde \Sigma _\theta ^{vec} - {{\rm{B}}_\theta }{{\bar {\cal P}}_\theta }} \right\|  \Gamma _\theta$ to the right-hand side and some manipulations yields 
\begin{align}
\sup \dot V(\tilde \Sigma _\theta ^{vec}) &=  - 2{\alpha _\theta } {{{\bf{A}}_\theta }} \left( { {{{\bf{A}}_\theta }} \tilde \Sigma _\theta ^{vec} - {{\rm{B}}_\theta }{{\bar {\cal P}}_\theta } } \right)   \left\| { {{{\bf{A}}_\theta }} \tilde \Sigma _\theta ^{vec} - {{\rm{B}}_\theta }{{\bar {\cal P}}_\theta }} \right\|  \Gamma _\theta \nonumber \\
 &\quad - 2{\alpha _\theta } {{{\bf{A}}_\theta }} {{{\rm{B}}_\theta }{{\bar {\cal P}}_\theta }} \left\| { {{{\bf{A}}_\theta }} \tilde \Sigma _\theta ^{vec} - {{\rm{B}}_\theta }{{\bar {\cal P}}_\theta }} \right\|  \Gamma _\theta \nonumber \\
& \le  - 2{\alpha _\theta } \left\| {{{\bf{A}}_\theta }\tilde \Sigma _\theta ^{vec} - {\rm{B}} _\theta \bar {\cal P}_\theta} \right\| + 2{\alpha _\theta }\frac{{{\lambda _{\max }}({{[{{\bf{A}}_\theta }]}^r})}}{{{\lambda _{\min }}({{[{{\bf{A}}_\theta }]}^{r + 1}})}}\left\| { {{{\bf{A}}_\theta }}  {{{\rm{B}}_\theta }{{\bar {\cal P}}_\theta }} } \right\|  \nonumber \\
&  \le  - 2{\alpha _\theta }\left\| {{{\bf{A}}_\theta }\tilde \Sigma _\theta ^{vec}} \right\| + \varepsilon (\tilde \Sigma _\theta ^{vec},{\bar {\cal P}_\theta }) \nonumber \\
&  \le  - 2{\alpha _\theta }{\left( {V(\tilde \Sigma _\theta ^{vec})} \right)^{\frac{1}{2}}} + \varepsilon (\tilde \Sigma _\theta ^{vec},{{\bar {\cal P}}_\theta }),
\label{eq:48}
\end{align}
where
\begin{align}
& \varepsilon (\tilde \Sigma _\theta ^{vec},{\bar {\cal P}_\theta }) =   2{\alpha _\theta }\big[\left\| {{{\rm{B}}_\theta }{{\bar {\cal P}}_\theta }} \right\| +   \frac{{{\lambda _{\max }}({{[{{\bf{A}}_\theta }]}^r})}}{{{\lambda _{\min }}({{[{{\bf{A}}_\theta }]}^{r + 1}})}}\left\| { {{{\bf{A}}_\theta }}  {{{\rm{B}}_\theta }{{\bar {\cal P}}_\theta }} } \right\|\big]. \label{eq:49}
\end{align}
Note that
\begin{align}
- 2{\alpha _\theta } \left\| {{{\bf{A}}_\theta }\tilde \Sigma _\theta ^{vec} - {\rm{B}} _\theta \bar {\cal P}_\theta} \right\| \le
  - 2{\alpha _\theta }\big[\left\|{{\bf{A}}_\theta }\tilde \Sigma _\theta ^{vec}\right\| - \left\|{{\rm{B}}_\theta }{\bar {\cal P}_\theta }\right\|\big], \label{eq:50}
\end{align}
and
\begin{align}
& \left\| {2{\alpha _\theta }{{{\bf{A}}_\theta }}  {{{\rm{B}}_\theta }{{\bar {\cal P}}_\theta }} } \right\| \left\| {{{\bf{A}}_\theta }\tilde \Sigma _\theta ^{vec} - {{\rm{B}}_\theta }{{\bar {\cal P}}_\theta }\,} \right\|  \|\Gamma _\theta  \| \nonumber \\
&  = 2{\alpha _\theta }\left\| {{{\bf{A}}_\theta }\tilde \Sigma _\theta ^{vec} - {{\rm{B}}_\theta }{{\bar {\cal P}}_\theta }\,} \right\|  \frac{{\left\| { {{{\bf{A}}_\theta }}  {{{\rm{B}}_\theta }{{\bar {\cal P}}_\theta }} } \right\|\left\| {{{\left[ {{{\bf{A}}_\theta }} \right]}^r}\left[ {{{{\bf{A}}_\theta }} \tilde \Sigma _\theta ^{vec} - {{\rm{B}}_\theta }{{\bar {\cal P}}_\theta }} \right]} \right\|}}{{{{\left[ { {{{\bf{A}}_\theta }} \tilde \Sigma _\theta ^{vec} - {{\rm{B}}_\theta }{{\bar {\cal P}}_\theta }} \right]}^{\top}}{{\left[ {{{\bf{A}}_\theta }} \right]}^{r + 1}}\left[ {{{{\bf{A}}_\theta }} \tilde \Sigma _\theta ^{vec} - {{\rm{B}}_\theta }{{\bar {\cal P}}_\theta }} \right]}} \nonumber \\
&  \le 2{\alpha _\theta }\left\| {{{\bf{A}}_\theta }\tilde \Sigma _\theta ^{vec} - {{\rm{B}}_\theta }{{\bar {\cal P}}_\theta }} \right\|\frac{{{\lambda _{\max }}({{[{{\bf{A}}_\theta }]}^r})\left\| {{{{\bf{A}}_\theta }}  {{{\rm{B}}_\theta }{{\bar {\cal P}}_\theta }}} \right\|\left\| {\left[ { {{{\bf{A}}_\theta }}\tilde \Sigma _\theta ^{vec} - {{\rm{B}}_\theta }{{\bar {\cal P}}_\theta }} \right]} \right\|}}{{{\lambda _{\min }}({{[{{\bf{A}}_\theta }]}^{r + 1}}){{\left\| {\left[ { {{{\bf{A}}_\theta }} \tilde \Sigma _\theta ^{vec} - {{\rm{B}}_\theta }{{\bar {\cal P}}_\theta }} \right]} \right\|}^2}}} \nonumber \\
&  \le 2{\alpha _\theta }\frac{{{\lambda _{\max }}({{[{{\bf{A}}_\theta }]}^r})}}{{{\lambda _{\min }}({{[{{\bf{A}}_\theta }]}^{r + 1}})}}\left\| { {{{\bf{A}}_\theta }}  {{{\rm{B}}_\theta }{{\bar {\cal P}}_\theta }} } \right\|.
\label{eq:51}
\end{align}}
Now, if the following inequalities hold
\[\tilde \Sigma _\theta ^{vec}\, > \frac{1}{{2{\alpha _\theta }}}\varepsilon (\tilde \Sigma _\theta ^{vec},{\bar {\cal P}_\theta })({{\cal H}_\theta }(t){{\cal H}_\theta }^{\top}(t) + \sum\limits_{j = 1}^{{p_\theta }} {{{\cal H}_\theta }({t_j}){{\cal H}_\theta }^{\top}({t_j}))}, \] 
then, $\sup \dot V(\tilde \Sigma _\theta ^{vec}) < 0$, which completes the proof.  \hfill $\square$

\subsection{Meta Learner}
To learn a Koopman operator model with minimum approximation error and dimension, a meta-learner is developed in this subsection. Toward this aim, a meta cost is defined as
\begin{align}
{J^R}(\Xi ): =  { {{\ell _\theta } ( {\hat A_\theta ,\hat B_\theta ,{\cal D}^{{\rm{eval }}}} )} } + \lambda {n_{{\xi _\theta }}}, \label{eq:56}
\end{align}   
{
 where $\Xi=[\theta(1),...,\theta({n_{{\xi _\theta }}})]^{\top}$ denotes the vector of the ordered set $\theta$,} and leads to the $(\hat A_\theta ,\hat B_\theta)$ model by the base learner layer, $\lambda$ is a constant, and ${n_{{\xi _\theta }}} = |\theta|$  promotes sparsity to find a set of observable with minimum cardinality. The meta loss function ${\ell_\theta }$ is the average identification error for a set of observables with respect to the collected data set ${\cal D}_{}^{eval{\rm{ }}}$ after the lower layer convergences to $(\hat A_\theta ,\hat B_\theta)$ model, and, it is defined as
\begin{align}
&{\ell _\theta }\left( {{{\hat A}_\theta },{{\hat B}_\theta },{{\cal D}^{{\rm{eval}}}}} \right) = \frac{1}{{{p_\theta }}}\sum\limits_{k = 1}^{{p_\theta }} {\| {{{\hat {\bar \xi} }_\theta }({{\cal D}^{{\rm{eval}}}},{{\hat A}_\theta },{{\hat B}_\theta },t,{\mkern 1mu} {t_k}) - {{\bar \xi }_\theta }({{\cal D}^{{\rm{eval}}}},{t_k})} \|}  \label{eq:57}
\end{align}
with
\begin{align}
\left\{ \begin{array}{l}
{{\hat {\bar \xi} }_{\theta }}(t) = {{\hat \Sigma }_{\theta i}}^{\top}{{\bar h}_{\theta }}(t,{\cal D}^{{\rm{eval }}}) + {{\cal A}_{\theta }}{{\bar l}_{\theta }}(t,{\cal D}^{{\rm{eval }}})\\
\,\,\,\,\,\,\,\,\,\,\,\,\,\,\,\,\,\,\,\,\,\quad  + {e^{ - {{\cal A}_{\theta }}t}}{{\bar \xi }_{\theta }}(t,{\cal D}^{{\rm{eval }}},0)\\
{{\dot h}_{\theta }}(t) =  - {a_{\theta }}{h_{\theta i}}(t,{\cal D}^{{\rm{eval }}}) + {{\cal Z}_{\theta i}}({\xi _{\theta }}(t,{\cal D}^{{\rm{eval }}}),{u})\\
{{\dot l}_{\theta }}(t) =  - {{\cal A}_{\theta }}{l_{\theta }}(t,{\cal D}_i^{{\rm{eval }}}) + {\xi _{\theta }}(t,{\cal D}^{{\rm{eval }}})
\end{array} \right. \label{eq:58}
\end{align} 
{
	The collected data set ${\cal D}^{eval}$ is formed by samples that are stored in the memory and represent the state space well. Since the proposed novel batch-online learning in the lower layer allows learning in finite time,  data set ${\cal D}^{eval}$ can provide a fair and control-agnostic comparison for sets of observables. It means that the Koopman parameters for selected observables by the meta layer can be computed in finite time without requiring additional rich incremental data. That is, different matrix pairs $(\hat A_\theta ,\hat B_\theta)$ are obtained in finite time by the lower layer learner as the set of observables varies, which result in different identifier system \eqref{eq:58}. 
 }
Therefore, ${J^R}(\Xi )$ is a function of sets $\theta$ which are parameterizing the lifted state vector of the base learner, i.e., ${\xi _\theta }(t)$. 

\begin{algorithm}[t]
\caption{Meta Learner}
\SetAlgoLined
\KwResult{Optimal observables vector ${\Xi ^*}$.}
\,\,\,\textbf{1:}{\,\, Evaluate meta-loss function: $J_{}^R \leftarrow \left\{ {J_{}^R(\Xi ),{\cal D}^{eval{\rm{ }}}} \right\}$ };  \\
\textbf{2:}{\,\, Assemble and construct a training dataset: ${\cal D} \leftarrow \left\{ {\Xi ;J_{}^R} \right\}$ };  \\
\textbf{3:}{\,\, Place a GP prior on $J_{}^R$};  \\ 
\textbf{4:}{\,\, Initialize the GP using ${\cal D}$} ; \\
\For{$k = 1$ to ${T_{outer}}$}{
\,\,\,\textbf{5.1:} \,  Train the GP that approximates ${J^R}$ on the basis of data ${\cal D}$;\\
\textbf{5.2:} \, Determine the acquisition function $\alpha (\Xi \mid {\cal D})$ using the GP; \\
\textbf{5.3:} \, Compute the subsequent vector $\Xi$ as
\begin{align}
\begin{array}{l}
{\Xi _{k + 1}} \leftarrow \mathop {\arg \max }\limits_{\Xi \in {\cal M}} \alpha (\Xi \mid {\cal D})\\
{\Xi _{k + 1}} = round({\Xi _{k + 1}})
\end{array} \label{eq:61}
\end{align}
where  ${\cal M}$ is some design space of interest; \\
\textbf{5.4:} \, Evaluate $J_{k + 1}^R$ based on ${\cal D}_{}^{{\rm{eval }}}$, and append it into ${\cal D}$:\,\,  ${\cal D} \leftarrow {\cal D} \cup \left\{ {{\Xi _{k + 1}};J_{k + 1}^R} \right\}$;     \\
\textbf{5.5:} \, 
If the stopping criterion is satisfied, exit the loop;
} 
\textbf{6:}{\,\, Return ${\Xi ^*} = round({\Xi _{{k^*}}})$ where
\begin{align}
{k^*} = \mathop {\arg \min }\limits_k J_k^R \label{eq:62}
\end{align}
} 
\end{algorithm}

 As delineated in Algorithm 1, to optimize the meta cost, a Bayesian optimization (BO) strategy for discrete variables \citep{shahriari2015taking,luong2019bayesian} is leveraged which leverages the previously recorded data sets in the memory, i.e., ${\cal D}^{{\rm{eval }}}$, to map the so-called meta-parameters $\theta $  to meta cost ${J^R}(\Xi)$. In Steps 1-2, the algorithm is started up by using the available collecting data samples ${\cal D}$ and ${\cal D}_{}^{{\rm{eval }}}$ for ${T_{outer}}$ randomly chosen different vectors of ${\Xi _k}$ (with $k = 1, \ldots ,{T_{outer}}$). The training samples are used by the lower layer learner update law and can be a combination of the samples collected incrementally as well as the memory samples. For each vector ${\theta _k}$, the meta cost, however, $J_k^R$ is evaluated using \eqref{eq:57}-\eqref{eq:58} and leveraging only the samples in the memory (to make the identifier control agnostic). As a result, the initial set ${\cal D} = \left\{ {({\Xi _k},J_k^R):k = 1,...,{T_{out}}} \right\}$ of libraries of observable functions and corresponding performance $J_{}^R$ is constructed. Since the function $J_{}^R$  is not known a priori, nonparametric Gaussian Process (GP) models should be used (Steps 3-4) to approximate it using the training dataset ${\cal D}$.  Therefore, GP model characterizing “the best guess” of $J_{}^R$ corresponding to the library of observable functions $\theta $, i.e., ${\cal L}(\theta )$, based on the available training dataset ${\cal D}$.  To this aim, 
 the function values $J_{}^R$, which are related to different sets of $\theta $, are assumed as random variables with a joint Gaussian distribution 
  (i.e.,  $J_{}^R$ is a Gaussian variable) dependent on the vector of $\Xi $  with the defined a prior mean function and variance
\begin{align}
& {m_k}(\Xi ) = {\bf{k}}_k^\prime {({{\bf{K}}_k} + \sigma _e^2I)^{ - 1}}J_k^R(\Xi ), \label{eq:63} \\
& \sigma _k^2(\Xi ) = \kappa (\Xi ,\Xi ) - {\bf{k}}_k^\prime {({{\bf{K}}_k} + \sigma _e^2I)^{ - 1}}{{\bf{k}}_k} + \sigma _e^2, \label{eq:64}
\end{align}
where $\sigma _e^2$  is the variance of Gaussian noise, and $\kappa (\Xi ,{\Xi _j})$ and 
$\kappa ({\Xi _j},{\Xi _m})$ denote the $j$-th and $[j,m]$-th entry of the vector ${{\bf{k}}_k}$ and  Kernel matrix ${{\bf{K}}_k}$, respectively. 
The covariance function $\kappa (\Xi ,\tilde \Xi )$ determines the covariance between $J_{}^R(\theta )$ and $J_{}^R(\tilde \theta )$ and is described as 
\begin{align}
\kappa (\Xi ,\tilde \Xi ) = \sigma _0^2{e^{ - \frac{1}{{2{\lambda ^2}}}\left[ {{\Xi ^\prime } - {{\tilde \Xi }^\prime }{\mu ^\prime } - {{\tilde \mu }^\prime }} \right]{{\left[ {{\Xi ^\prime } - {{\tilde \Xi }^\prime }{\mu ^\prime } - {{\tilde \mu }^\prime }} \right]}^\prime }}}, \label{eq:65}
\end{align}
where ${\sigma _0}$ and $\lambda $ are the design parameters.

Afterward, the algorithm is repeated till a predefined termination criterion is satisfied.
 Following are the steps that are performed during each iteration.
  Based on the available training dataset ${\cal D}$, a GP is fitted in Step 5.1.
   To find the next library of observable functions ${\theta _{k + 1}}$, the function $\alpha (\Xi \mid {\cal D})$ (which is called the acquisition function) is getting optimized in Step 5.2. 
    This function $\alpha (.)$ is determined using the estimated GP mean and covariance and its main objective is to balance between exploration and exploitation. 
That is, exploring by evaluating the function ${J^R}$ in domains of the search space with high variance and also exploiting the past recorded data and optimizing the expected improvement over domains with high mean.
      Now, let the acquisition function $\alpha (.)$ be defined as
\begin{align}
	\alpha (\Xi \mid {\cal D})  = \mathbb{E} \left[ {\max \left\{ {0,J_{}^{R*} - {J^R}(\Xi )} \right\}} \right], \label{eq:66}
\end{align}
where the target value $J_{}^{R*}$  is the minimum of all explored data, represents the most optimal value of $J_{}^{R}$, and defined as
\begin{align}
J_{}^{R*} = \mathop {\arg \min }\limits_k J_k^R. \label{eq:67}
\end{align}
The acquisition function given in \eqref{eq:66} can be computed analytically based on the mentioned GP framework as
\begin{align}
\alpha (\Xi \mid {\cal D})  =  \left\{ {\begin{array}{*{20}{c}}
{\left( {J_{}^{R*} - {m_k}(\Xi )} \right)\Phi (Z) + {\sigma _k}(\Xi )\psi (Z)}&{{\sigma _i}(\Xi ,v) > 0}\\
0&{O.W.}
\end{array}} \right. \label{eq:68}
\end{align}
where $Z = {{(J_{}^{R*} - {m_k}(\Xi ))} \mathord{\left/
 {\vphantom {{(J_{}^{R*} - {m_k}(\Xi ))} {{\sigma _k}(\Xi )}}} \right.
 \kern-\nulldelimiterspace} {{\sigma _k}(\Xi )}}$,  $\Phi $ denotes the cumulative density function, and $\psi $  is the probability density function. 
 Intuitively, the acquisition function given in \eqref{eq:66}   selects the succeeding parameter point at which the improvement over  $J_{}^{R*}$ should be the most in expectation.
   Note that it is no need for real physical interaction with the system in order to optimize $\alpha (\Xi \mid {\cal D})$ in \eqref{eq:61}, and only need to evaluate the GP model.
    If any new data sets ${\cal D}_{k + 1}$ and ${\cal D}_{k + 1}^{{\rm{eval }}}$ are available, they will be collected and augmented to the available data sets ${\cal D}$  and ${\cal D}_{}^{{\rm{eval }}}$  in Step 5.4. In Step 5.5, the new determined optimal vector ${\Xi ^*}$ is evaluated based on the updated evaluation data of real the system. 

\begin{rem}\label{Remark:5}
For the sake of simplicity, the BO with naive rounding (Naive BO) approach is used to deal with the discrete nature of $\Xi $. That is, we treat discrete variables as continuous then apply a normal BO method, and finally rounds the suggested continuous point to the nearest discrete point before function evaluations. 
\end{rem}

\begin{rem}\label{Remark:6}
Using Bayesian optimization, bounds can be set on the search space of $\theta $.
When maximization of the acquisition function is performed in Algorithm 1, then, these bounds can be incorporated into. 
  In general, when the search space is narrowed down, the algorithm tends to converge faster, therefore,  requiring fewer time-consuming evaluations of the functional ${J^R}(\Xi )$. 
  Specifically, each evaluation of the functional ${J^R}(\Xi )$ takes 
\begin{align}
{t_{\theta }}^ \star  = \frac{1}{{{\alpha _{\theta }}}}\| {{\bf{H}}_\theta(0)} \|. \label{eq:69}
\end{align}
Prior system knowledge and design choices may be exploited to define suitable bounds.  Moreover, the proposed method is agnostic to the choice of the controller $u$ since it relies on collected data in the memory to evaluate the meta cost. However, if we have some prior knowledge of the controller $u$,
it would also be leveraged to narrow the search space and consequently accelerate the convergence of the algorithm.
\end{rem}


{
\begin{rem}\label{Remark:9}
	It is worth noting that a proposed batch learning method developed by \citep{lehrer2010parameter}, aims to identify uncertain discrete-time systems in finite-time, requiring the online invertibility check of regressor matrix and its inverse computation as well as interval excitation of regressor. In the case of large numbers of unknown parameters, however, the required inversion of the regressor matrix makes the method in \citep{lehrer2010parameter} computationally inefficient for online learning.
\end{rem}}

\section{Simulation}
In this section, two simulation results are presented to validate the theoretical results. 

{\bf{Example 1:}} 
The following nonlinear system is considered as the system dynamics
\begin{align}
{\begin{array}{*{20}{l}}
{{{\dot x}_1} = \mu {x_1}}\\
{{{\dot x}_2} = \lambda \left( {{x_2} - x_1^4 + 2x_1^2} \right) + u}
\end{array}} \label{eq:70}
\end{align}
where it can be rewritten as 
\begin{align}
\begin{array}{l}
\frac{d}{{dt}}\underbrace {\left[ {\begin{array}{*{20}{l}}
{{y_1}}\\
{{y_2}}\\
{{y_3}}\\
{{y_4}}
\end{array}} \right]}_\xi  = \underbrace {\left[ {\begin{array}{*{20}{c}}
\mu &0&0&0\\
0&\lambda &{2\lambda }&{ - \lambda }\\
0&0&{2\mu }&0\\
0&0&0&{4\mu }
\end{array}} \right]}_{{A^*}}\underbrace {\left[ {\begin{array}{*{20}{l}}
{{y_1}}\\
{{y_2}}\\
{{y_3}}\\
{{y_4}}
\end{array}} \right]}_\xi  + \underbrace {\left[ {\begin{array}{*{20}{l}}
0\\
1\\
0\\
0
\end{array}} \right]}_{{B^*}}u,
\end{array} \label{eq:71}
\end{align}
with $\xi  = {[{x_1},{x_2},x_1^2,x_1^4]^{\top}}$, $\mu  =  - 1$, $\lambda  = -1$.

To satisfy Condition \ref{cond:3}, we inject the following probing noise into the control input (i.e., add it to the control input $u$)  for $t \in (0~-~0.5)~sec $ 
{
\begin{align}
\begin{array}{l}
\begin{array}{l}
	1.25e^{-t}(0.4{(sin(0.1t))^6}cos(1.5t) + 0.3{(sin(2.3t))^4}cos(0.7t) + 0.5cos(2.4t){(sin(7.4t))^2}\\
	+ 0.4{(sin(2.6t))^5} + 0.7{(sin(3t))^2}cos(4t) + 0.3sin(0.3t){(cos(1.2t))^2} + 0.4{(sin(1.12t))^3}\\
	+ 0.3{(sin(4t))^3} + 0.4{(sin(3.5t))^5} + 0.4cos(2t){(sin(5t))^8} + 0.3sin(t){(cos(0.8t))^2}\\
	+ 0.5{(cos(2.4t))^3}{(sin(7.4t))^2} + 0.1{(sin(3.5t))^7} + 0.1{(cos(2t))^4}{(sin(5t))^4} + 0.4{(sin(2.1t))^2}\\
	+ 0.3{(sin(2.1t))^3}{(cos(0.9t))^3} + 0.1{(sin(1.7t))^5}{(cos(0.9t))^2} + 0.1{(sin(0.4t))^2}{(cos(1.6t))^3})
\end{array}
\end{array} \label{eq:73j}
\end{align}}

The proposed learning scheme is used to find the best library of observable functions $\theta $, and matrices ${A^*}$, and ${B^*}$. 
In the history stack, the size of recorded data is set as $21$.

We choose a  finite set of polynomial observable functions based on the monomials of  states $x_1$ and $x_2$ as follows
\begin{align}
{\cal C}(k)=\left\{x_{1}^{a} \cdot x_{2}^{b}  \mid a, b \in\{0,1,2, \ldots k\}\right\}  \cup \{ x_1^4\},  \label{eq:73}
\end{align}
{where $k=2$ is the order of the basis functions,
 ${\xi _{1}}(x)=x_1$, ${\xi _{2}}(x)=x_2$, ${\xi _{3}}(x)=x_1 \cdot x_2$, ${\xi _{4}}(x)=x_1^2$, ${\xi _{5}}(x)=x_2^2$, ${\xi _{6}}(x)=x_1^2 \cdot x_2$, ${\xi _{7}}(x)=x_1 \cdot x_2^2$,  ${\xi _{8}}(x)=x_1^2 \cdot x_2^2$, and ${\xi _{9}}(x) = x_1^4$. Now, for an ordered set $\theta$ where $\theta \subseteq  \{1,...,9\}$ and ${n_{{\xi _\theta }}}= |\theta| \le 9$,  the vector ${\xi _\theta }(t) = {[ {{\xi _{\theta (1)}}(x(t)), \cdots, {\xi _{\theta({n_{{\xi _\theta }}})}}(x(t))} ]^{\top}}$ can be defined, by using Definition \ref{defn:5}, to lift the system from a state-space to function space of observables. For instance,  the ordered set $\theta=\{ 1,2,4,9\}$ corresponds with the lifted state vector ${\xi _{\theta=\{ 1,2,4,9\}}}(t) = [{\xi _{1}}(x)=x_1,~  {\xi _{2}}(x)=x_2,~ {\xi _{4}}(x)=x_1^2,~ {\xi _{9}}(x) = x_1^4]^{\top}$.}



\begin{figure}[!t]
\centering{\includegraphics [width=3.5in] {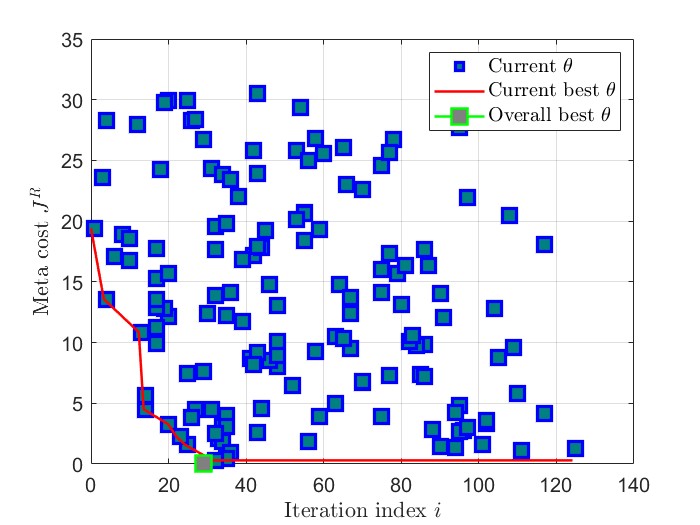}}
\caption{Meta cost $J^R$ vs iteration $i$ of Algorithm 1. At each iteration $i$, the cost of the current set of observable functions $\theta$ and of the current best one are depicted. The green square shows the overall best $\theta=\{ 1,2,4,9\}$ which is found at iteration $29$.}  \label{FigBys1}
\end{figure}

\begin{figure}[!t]
\centering{\includegraphics [width=3.5in] {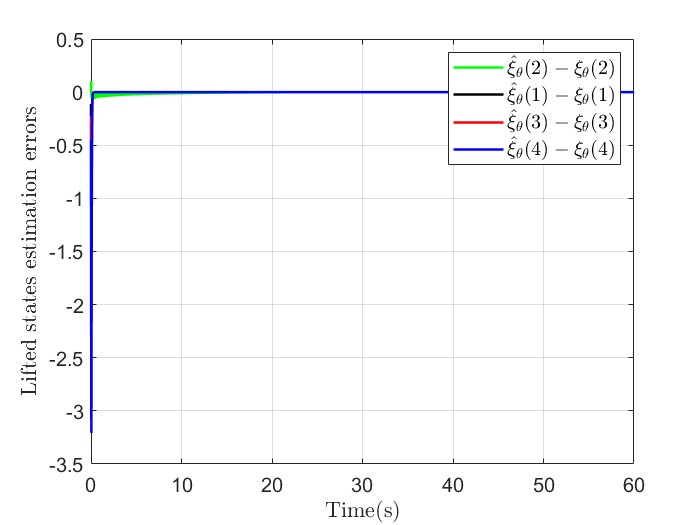}}
\caption{Lifted state estimation errors with the library of observables $\theta =  \{1,2,4,9\}$.}  \label{Fig10CL}
\end{figure}

\begin{figure}[!t]
\centering{\includegraphics [width=3.5in] {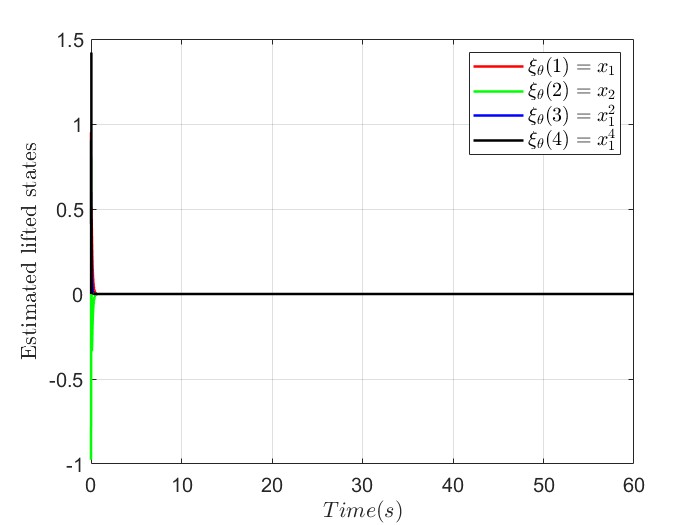}}
\caption{Lifted state trajectories with the library of observables $\theta =  \{1,2,4,9\}$.}  \label{Fig3CL}
\end{figure}

\begin{figure}[!t]
\centering{\includegraphics [width=3.5in] {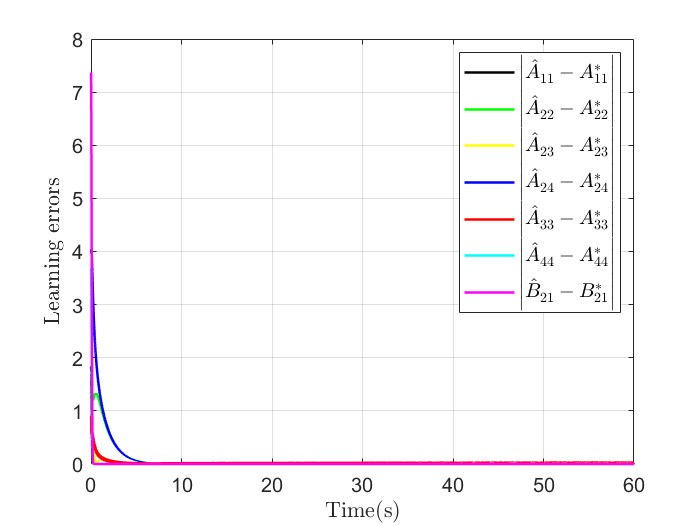}}
\caption{Online learning error with the library of observables $\theta =  \{1,2,4,9\}$.}  \label{Fig8CL}
\end{figure}

\begin{figure}[!t]
\centering{\includegraphics [width=3.5in] {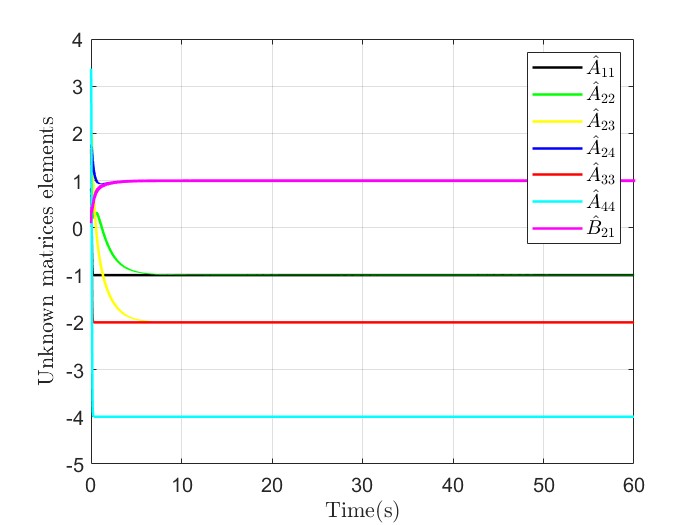}}
\caption{Estimated unknown elements of $A^*$ and $B^*$ with the library of observables $\theta =  \{1,2,4,9\}$. Note that for simplicity of demonstration, we only show a non-zero elements of matrices. } \label{Fig2CL}
\end{figure}

The system is initialized at $x_1=1$ and $x_2=-1$. Let $\alpha_\theta = 3$. 
MATLAB Statistics and Machine Learning Toolbox is used to implement Algorithm 1, with the EI in \eqref{eq:66} serving as an acquisition function.
 The meta cost $J^R$ vs iteration $i$ of Algorithm 1 is illustrated in Fig.~\ref{FigBys1}. Fig.~\ref{Fig10CL} shows that,  within finite time, the lifted state estimation error of the base learner for $\theta  = \{ 1,2,4,9\} $ has been zeroed out. The lifted states trajectories evolution is described in Fig.~\ref{Fig3CL}. The convergence of the base learner parameters
 {
\[{\hat A_{\theta  = \{ 1,2,4,9\} }} = \left[ {\begin{array}{*{20}{c}}
		{{\rm{ - 1}}}&0&{\rm{0}}&0\\
		{\rm{0}}&{{\rm{ - }}1}&{{\rm{ - }}2}&{\rm{1}}\\
		{\rm{0}}&0&{{\rm{ - 2}}}&0\\
		{\rm{0}}&{\rm{0}}&0&{{\rm{ - 4}}}
\end{array}} \right],\,\,{\hat B_{\theta  = \{ 1,2,4,9\} }} = {\left[ {\begin{array}{*{20}{c}}
			{\rm{0}}&{\rm{1}}&0&{\rm{0}}
	\end{array}} \right]^ \top },\]}
to the true values $A^*$ and $B^*$ within finite time and during online learning has been shown in Fig.~\ref{Fig8CL} and Fig.~\ref{Fig2CL} .

\begin{figure}[!t]
\centering{\includegraphics [width=3.5in] {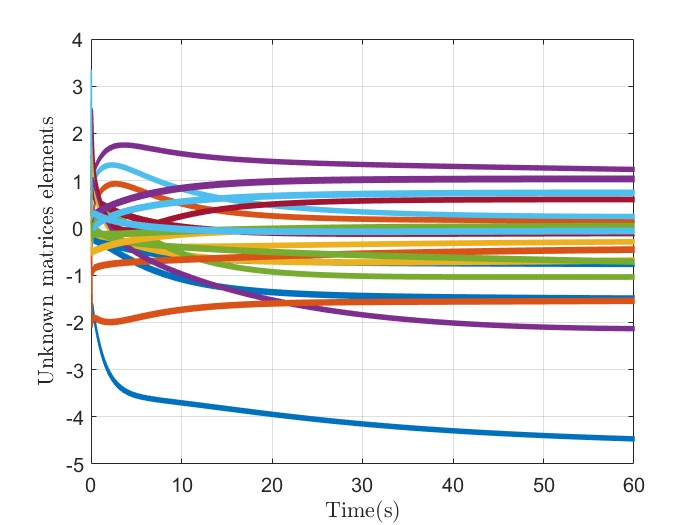}}
\caption{Estimated unknown elements of $A^*$ and $B^*$ with the library of observables $\theta =  \{1,2,4,8\}$.} \label{Fig6CL}
\end{figure}

For the case of $\theta =  \{1,2,4,8\}$, the finite time convergence of the base learner parameters 
\[\begin{array}{l}
{{\hat A}_{\theta  = \{ 1,2,4,8\} }} = \left[ {\begin{array}{*{20}{c}}
{{\rm{ - 0}}{\rm{.8316}}}&{{\rm{0}}{\rm{.2151}}}&{{\rm{1}}{\rm{.1806}}}&{{\rm{ - 0}}{\rm{.7588}}}\\
{{\rm{ - 0}}{\rm{.1784}}}&{{\rm{ - 0}}{\rm{.9781}}}&{{\rm{  0}}{\rm{.1839}}}&{{\rm{ -1}}{\rm{.5642}}}\\
{{\rm{ - 0}}{\rm{.3880}}}&{{\rm{ - 0}}{\rm{.3554}}}&{{\rm{ - 2}}{\rm{.0771}}}&{{\rm{ - 0}}{\rm{.6459}}}\\
{{\rm{0}}{\rm{.6716}}}&{{\rm{0}}{\rm{.6625}}}&{{\rm{ - 4}}{\rm{.5250}}}&{{\rm{ - 1}}{\rm{.6133}}}
\end{array}} \right]
\end{array}\]
\[{{\hat B}_{\theta  = \{ 1,2,4,8\} }} = {\left[ {\begin{array}{*{20}{c}}
{{\rm{ - 0}}{\rm{.0973}}}&{{\rm{0}}{\rm{.9653}}}&{{\rm{  0}}{\rm{.1077}}}&{{\rm{ - 0}}{\rm{.1402}}}
\end{array}} \right]^{\top}}\]
to neighborhood values of the true matrices $A^*$ and $B^*$ during online learning has been shown in Fig.~\ref{Fig6CL}

{\bf{Example 2:}} 

\begin{figure}[!t]
	\centering{\includegraphics [width=3in] {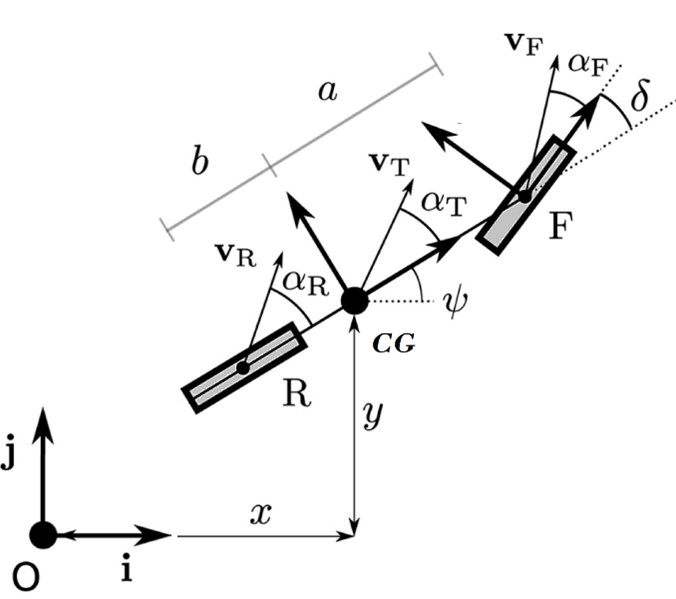}}
	\caption{The single-track model. 
		In this simulation, all tires on the same axle are assumed to have equal slip angles, and each axle is represented by a single tire with equivalent dynamic properties.} 
	\label{fig1e2}
\end{figure}

The vehicle is modeled as a nonlinear single-track bicycle model (See Fig.~\ref{fig1e2}). 
The yaw dynamics of a simple vehicle can be qualitatively described and analyzed with this model in all operating conditions. 
Using the Lagrangian approach, the state equation of this vehicle model can be written as \citep{andrzejewski2006nonlinear,de2016vehicle}
\begin{align}
\Scale[0.92]{\begin{array}{l}
{{\dot v}_T} = \left( {{F_{x,\mathbf{F}}}\cos \left( {{\alpha _T} - \delta } \right) + {F_{x,\mathbf{R}}}\cos \left( {{\alpha _T}} \right)} \right)/{m_T} + \left( {{F_{y,\mathbf{F}}}\sin \left( {{\alpha _T} - \delta } \right) + {F_{y,\mathbf{R}}}\sin \left( {{\alpha _T}} \right)} \right)/{m_T}\\
{{\dot \alpha }_T} = \left( { - {F_{x,\mathbf{F}}}\sin \left( {{\alpha _T} - \delta } \right) - {F_{x,\mathbf{R}}}\sin \left( {{\alpha _T}} \right)} \right)/\left( {{m_T}{v_T}} \right) + \left( {{F_{y,\mathbf{F}}}\cos \left( {{\alpha _T} - \delta } \right) + {F_{y,\mathbf{R}}}\cos \left( {{\alpha _T}} \right) - {m_T}{v_T}\dot \psi } \right)/\left( {{m_T}{v_T}} \right)\\
\ddot \psi  = \left( {{F_{x,\mathbf{F}}}a\sin (\delta ) + {F_{y,\mathbf{F}}}a\cos (\delta ) - {F_{y,\mathbf{R}}}b} \right)/{I_T}
\end{array} }\label{eq:6e2}
\end{align}
where the longitudinal forces acting on the front and rear tires, respectively, are given by  
 ${F_{x,\mathbf{F}}}$  and  ${F_{x,\mathbf{R}}}$. Moreover,  
   ${F_{y,\mathbf{F}}}$  and ${F_{y,\mathbf{R}}}$ (the lateral forces) are also determined by the selected tire model where the subscripts $_\mathbf{F}$  and $_\mathbf{R}$ are denotes the front and rear points to which theses forces are associated. Moreover, ${\alpha _T}$ denotes the slip angle, and ${v_T}$  denotes the vehicle center of gravity velocity, 
   ${m_T}$  denotes the vehicle mass,  $\delta $ denotes the front axle steering angle,  ${I_T}$ is the vehicle inertia, and the distances between the points $\mathbf{F}$ , CG, and $\mathbf{R}$ are determined by the constants $a$  and $b$. 

{
The following equation describes tire model
\begin{align}
{{{F}}_{{y,\mathbf{F}}}} = K {{\alpha}_\mathbf{F}} ,  \nonumber \\
{{{F}}_{{y,\mathbf{R}}}} = K {{\alpha}_\mathbf{R}} ,
 \label{eq:7e2}
\end{align}
where the constant $K$ is stiffness, and
\begin{align}
	\begin{array}{l}
	{{\alpha}_\mathbf{F}} = {\alpha _{\rm{T}}} + \frac{a}{{{v_{{\rm{T}},0}}}}\dot \psi  - \delta, \\
	{{\alpha}_\mathbf{R}} = {\alpha _{\rm{T}}} - \frac{b}{{{v_{{\rm{T}},0}}}}\dot \psi .
	\end{array}  \label{eq:7e2slip}
\end{align}}


{
We first reformulate the vehicle dynamics \eqref{eq:6e2} into nonlinear affine in control system and then develop a novel hierarchical learning structure that leverages a novel incremental finite-time Koopman-based update law to approximately learn the linear representation of vehicle dynamics \eqref{eq:6e2}. To this aim, using the facts that 
\begin{align}
\begin{array}{l}
\cos ({\alpha _{\rm{T}}} - \delta ) = \cos {\alpha _{\rm{T}}}\cos \delta  + \sin {\alpha _{\rm{T}}}\sin \delta, \\
\sin ({\alpha _{\rm{T}}} - \delta ) = \sin {\alpha _{\rm{T}}}\cos \delta  - \cos {\alpha _{\rm{T}}}\sin \delta . \nonumber
\end{array}
\end{align}
One can rewrite \eqref{eq:6e2}  as
\begin{align}
\begin{array}{l}
	{{\dot v}_{\rm{T}}} = \frac{2}{{{m_T}}}K{\alpha _{\rm{T}}}^2 + \frac{1}{{{m_T}}}K\frac{{(a - b)}}{{{v_{{\rm{T}},0}}}}\dot \psi {\alpha _{\rm{T}}} - \frac{1}{{{m_T}}}K{\alpha _{\rm{T}}}\delta  - (\frac{1}{{2{m_T}}}K{\alpha _{\rm{T}}}^2 + \frac{1}{{2{m_T}}}K\frac{a}{{{v_{{\rm{T}},0}}}}\dot \psi {\alpha _{\rm{T}}}){\delta ^2} + \frac{1}{{2{m_T}}}K{\alpha _{\rm{T}}}{\delta ^3}\\
	\,\,\,\,\,\,\,\, + (\frac{1}{{{m_T}}}K\frac{a}{{{v_{{\rm{T}},0}}}}\dot \psi \frac{{{\alpha _{\rm{T}}}^2}}{2} - \frac{1}{{{m_T}}}K\frac{a}{{{v_{{\rm{T}},0}}}}\dot \psi  - \frac{1}{{{m_T}}}K{\alpha _{\rm{T}}} + \frac{1}{{{m_T}}}K{\alpha _{\rm{T}}}\frac{{{\alpha _{\rm{T}}}^2}}{2})\sin \delta  + (\frac{1}{{{m_T}}}K - \frac{1}{{{m_T}}}K\frac{{{\alpha _{\rm{T}}}^2}}{2})\delta \sin \delta \,\,\,\,\,\,\,\,\\
	{{\dot \alpha }_{\rm{T}}} = \frac{2}{{{m_T}{v_{\rm{T}}}}}K{\alpha _{\rm{T}}} + \frac{1}{{{m_T}{v_{\rm{T}}}}}K\frac{{(a - b)}}{{{v_{{\rm{T}},0}}}}\dot \psi  - \frac{1}{{{m_T}{v_{\rm{T}}}}}K\frac{{(a - b)}}{{{v_{{\rm{T}},0}}}}\dot \psi \frac{{{\alpha _{\rm{T}}}^2}}{2} - \frac{2}{{{m_T}{v_{\rm{T}}}}}K\frac{{{\alpha _{\rm{T}}}^3}}{2} - \dot \psi  + (\frac{1}{{{m_T}{v_{\rm{T}}}}}K\frac{{{\alpha _{\rm{T}}}^2}}{2} - \frac{1}{{{m_T}{v_{\rm{T}}}}}K)\delta \\
	\,\,\,\,\,\,\, + (\frac{1}{{2{m_T}{v_{\rm{T}}}}}K{\alpha _{\rm{T}}}\frac{{{\alpha _{\rm{T}}}^2}}{2} - \frac{1}{{2{m_T}{v_{\rm{T}}}}}K\frac{a}{{{v_{{\rm{T}},0}}}}\dot \psi  + \frac{1}{{2{m_T}{v_{\rm{T}}}}}K\frac{a}{{{v_{{\rm{T}},0}}}}\dot \psi \frac{{{\alpha _{\rm{T}}}^2}}{2} - \frac{1}{{2{m_T}{v_{\rm{T}}}}}K{\alpha _{\rm{T}}}){\delta ^2}\\
	\,\,\,\,\,\,\, + (\frac{1}{{2{m_T}{v_{\rm{T}}}}}K - \frac{1}{{{m_T}{v_{\rm{T}}}}}K\frac{{{\alpha _{\rm{T}}}^2}}{4}){\delta ^3} + (\frac{1}{{{m_T}{v_{\rm{T}}}}}K{\alpha _{\rm{T}}}^2 + \frac{1}{{{m_T}{v_{\rm{T}}}}}K\frac{a}{{{v_{{\rm{T}},0}}}}\dot \psi {\alpha _{\rm{T}}})\sin \delta  - \frac{1}{{{m_T}{v_{\rm{T}}}}}K{\alpha _{\rm{T}}}\delta \sin \delta \\
	\ddot \psi  = \frac{1}{{{I_T}}}K{\alpha _{\rm{T}}}(a - b) + \frac{1}{{{I_T}}}K\frac{{({a^2} + {b^2})}}{{{v_{{\rm{T}},0}}}}\dot \psi  - \frac{1}{{{I_T}}}K\delta a - (\frac{1}{{2{I_T}}}K\frac{{{a^2}}}{{{v_{{\rm{T}},0}}}}\dot \psi  + \frac{1}{{2{I_T}}}K{\alpha _{\rm{T}}}a){\delta ^2} + \frac{1}{{2{I_T}}}Ka{\delta ^3}
\end{array} \label{eq:8e2}
\end{align}}

Observing the facts that $\sin \theta  \approx \theta $  for $\theta  \le {13.99^o}$ and $\cos \theta  \approx 1 - {{{{(\theta )}^2}} \mathord{\left/
 {\vphantom {{{{(\theta )}^2}} 2}} \right.
 \kern-\nulldelimiterspace} 2}$  for $\theta  \le {37.93^\circ}$  and the facts that ${0^\circ} \le {\alpha _{\rm{T}}} \le {10^\circ}$  for race and high-performance tires and the number is a little lower for street tires, and assuming  $\left| \delta  \right| \le {35^\circ}$, and some manipulations, \eqref{eq:8e2} can be rewritten into the following affine in control form
{
\begin{align}
\begin{array}{l}
	\dot X = \left[ {\begin{array}{*{20}{c}}
			{{{\dot v}_{\rm{T}}}}\\
			{{{\dot \alpha }_{\rm{T}}}}\\
			{\ddot \psi }
	\end{array}} \right] = \underbrace {\left[ {\begin{array}{*{20}{c}}
				{\frac{2}{{{m_T}}}K{\alpha _{\rm{T}}}^2 + \frac{1}{{{m_T}}}K\frac{{(a - b)}}{{{v_{{\rm{T}},0}}}}\dot \psi {\alpha _{\rm{T}}}}\\
				{\frac{2}{{{m_T}{v_{\rm{T}}}}}K{\alpha _{\rm{T}}} + \frac{1}{{{m_T}{v_{\rm{T}}}}}K\frac{{(a - b)}}{{{v_{{\rm{T}},0}}}}\dot \psi  - \frac{1}{{{m_T}{v_{\rm{T}}}}}K\frac{{(a - b)}}{{{v_{{\rm{T}},0}}}}\dot \psi \frac{{{\alpha _{\rm{T}}}^2}}{2} - \frac{2}{{{m_T}{v_{\rm{T}}}}}K\frac{{{\alpha _{\rm{T}}}^3}}{2} - \dot \psi }\\
				{\frac{1}{{{I_T}}}K{\alpha _{\rm{T}}}(a - b) + \frac{1}{{{I_T}}}K\frac{{({a^2} + {b^2})}}{{{v_{{\rm{T}},0}}}}\dot \psi }
		\end{array}} \right]}_{{\cal I}(.)}\\
	+ \underbrace {\left[ {\begin{array}{*{20}{c}}
				{\begin{array}{*{20}{c}}
						{{{\cal G}_{11}}}\\
						{{{\cal G}_{21}}}\\
						{{{\cal G}_{31}}}
				\end{array}}&{\begin{array}{*{20}{c}}
						{{{\cal G}_{12}}}\\
						{{{\cal G}_{22}}}\\
						{{{\cal G}_{32}}}
				\end{array}}&{\begin{array}{*{20}{c}}
						{{{\cal G}_{13}}}\\
						{{{\cal G}_{23}}}\\
						{{{\cal G}_{33}}}
				\end{array}}&{\begin{array}{*{20}{c}}
						{\begin{array}{*{20}{c}}
								{{{\cal G}_{14}}}\\
								{{{\cal G}_{24}}}\\
								{{{\cal G}_{34}}}
						\end{array}}&{\begin{array}{*{20}{c}}
								{{{\cal G}_{15}}}\\
								{{{\cal G}_{25}}}\\
								{{{\cal G}_{35}}}
						\end{array}}
				\end{array}}
		\end{array}} \right]}_{{\cal G}(.)}\underbrace {\left[ {\begin{array}{*{20}{c}}
				\delta \\
				{\sin \delta }\\
				{\delta \sin \delta }\\
				{\begin{array}{*{20}{c}}
						{{\delta ^2}}\\
						{{\delta ^3}}
				\end{array}}
		\end{array}} \right]}_{{\cal U}(.)}
\end{array}
\label{eq:9e2}
\end{align}}
{where ${{{\cal G}_{11}} = \frac{{ - K{\alpha _{\rm{T}}}}}{{{m_T}}}}$, 
	${{{\cal G}_{12}} = \frac{{ - Ka}}{{{m_T}{v_{{\rm{T}},0}}}}\dot \psi  + \frac{{Ka{\alpha _{\rm{T}}}^2}}{{2{m_T}{v_{{\rm{T}},0}}}}\dot \psi  - \frac{{K{\alpha _{\rm{T}}}}}{{{m_T}}} + \frac{{K{\alpha _{\rm{T}}}^3}}{{2{m_T}}}}$,
	 ${{{\cal G}_{13}} = \frac{K}{{{m_T}}} - \frac{{K{\alpha _{\rm{T}}}^2}}{{2{m_T}}}}$, 
	 ${{{\cal G}_{14}} =   \frac{{-K{\alpha _{\rm{T}}}^2}}{{2{m_T}}} - \frac{{Ka{\alpha _{\rm{T}}}}}{{2{m_T}{v_{{\rm{T}},0}}}}\dot \psi }$,
 ${{{\cal G}_{15}} = \frac{{K{\alpha _{\rm{T}}}}}{{2{m_T}}}}$, ${{{\cal G}_{21}} = \frac{{ - K}}{{{m_T}{v_{\rm{T}}}}} + \frac{{K{\alpha _{\rm{T}}}^2}}{{2{m_T}{v_{\rm{T}}}}}}$, ${{{\cal G}_{22}} = \frac{{K{\alpha _{\rm{T}}}^2}}{{{m_T}{v_{\rm{T}}}}} + \frac{{Ka{\alpha _{\rm{T}}}}}{{{m_T}{v_{\rm{T}}}{v_{{\rm{T}},0}}}}\dot \psi }$, ${{{\cal G}_{23}} =  - \frac{{K{\alpha _{\rm{T}}}}}{{{m_T}{v_{\rm{T}}}}}}$,
 ${{\cal G}_{24}} = \frac{{K{\alpha _{\rm{T}}}^3}}{{4{m_T}{v_{\rm{T}}}}} - \frac{{Ka}}{{2{m_T}{v_{\rm{T}}}{v_{{\rm{T}},0}}}}\dot \psi  + \frac{{Ka{\alpha _{\rm{T}}}^2}}{{4{m_T}{v_{\rm{T}}}{v_{{\rm{T}},0}}}}\dot \psi  - \frac{{K{\alpha _{\rm{T}}}}}{{2{m_T}{v_{\rm{T}}}}}$, ${{{\cal G}_{25}} = \frac{K}{{2{m_T}{v_{\rm{T}}}}} - \frac{{K{\alpha _{\rm{T}}}^2}}{{4{m_T}{v_{\rm{T}}}}}}$, ${{{\cal G}_{31}} = \frac{{ - Ka}}{{{I_T}}}}$, ${{{\cal G}_{32}} = 0}$, ${{{\cal G}_{33}} = 0}$, ${{{\cal G}_{34}} = \frac{{ - K{a^2}}}{{2{I_T}{v_{{\rm{T}},0}}}}\dot \psi  - \frac{{K{\alpha _{\rm{T}}}a}}{{2{I_T}}}}$, and ${{{\cal G}_{35}} = \frac{{Ka}}{{2{I_T}}}}$. Moreover, ${\cal I}({\cal X})$,  ${\cal G}({\cal X})$, and ${\cal U}(\delta )$ are the drift dynamics of the system, the control input, and the input dynamics of the system, respectively.}


 The vehicle model for this study is a nonlinear single-track bicycle given in \eqref{eq:9e2} and shown in Fig.~\ref{fig1e2}. 
It is noteworthy that  this model is capable of qualitatively describing a simple nonlinear vehicle yaw dynamics in all operating conditions corresponding to $\left| {{\alpha _{\rm{T}}}} \right| \le 0.174533   \,\, rad$, $\left|\psi\right| \le 0.3 \,\, rad$, $\left|{\dot \psi} \right| \le 0.7 \,\, rad/{\mathop{\rm s}\nolimits} $, and $\left|\delta \right| \le {37.93}^\circ $.

\begin{table}[!t]
 \renewcommand{\arraystretch}{1.3}
 \caption{Vehicle parameters.}  \label{TB:2}
 \label{Table1}
 \centering
 \begin{tabular}{|c|c|c|}
 \hline
 {{\rm{Item}}} & Value & {{\rm{Description}}} \\
  \hline
 ${{m_T}}$ & {1300.0 {{\rm{kg}}}} & {{\rm{Total mass}}}  \\
 \hline
  ${{I_T}}$ & ${{{10}^4}} {{\rm{kg}} \cdot {{\rm{m}}^2}}$ & {{\rm{Moment of inertia}}}  \\
  \hline
  ${{K}}$ & $4\times{{10}^4}$ & {{\rm{Tires stiffness}}}  \\
   \hline
  $\alpha_{T, 0}$ & 0 rad  & {{\rm{\text { Initial vehicle sideslip angle }}}}  \\
  \hline
  $\dot{\Psi}_{0} $ & 0.01 $rad/s$ & {{\rm{\text {Initial yaw rate }}}}  \\
    \hline
  ${v}_{0} $ & 20 $m/s$ & {{\rm{\text {Initial velocity }}}}  \\
   \hline
    $a $ &  1.6154 &  Distance to CG \\
     \hline
     $b $ & 1.8846 &  Distance to CG \\
 \hline
 \end{tabular}
 \end{table}


 Assuming that $v_{\rm{T}} \ge 1 {m \mathord{\left/
 {\vphantom {m s}} \right.
 \kern-\nulldelimiterspace} s}$, and let the observable functions consist of the elements of the state vector $\cal{X}$ as:
\begin{align}
\Scale[1]{\begin{array}{l}
 {\cal C} = \{ \psi, {\alpha _T},\dot \psi ,{{\alpha _T}^2},{{\alpha _{\rm{T}}}^3},{{\alpha _{\rm{T}}}^4},{{\alpha _{\rm{T}}}^5},\frac{{{\alpha _T}}}{{{v_{\rm{T}}}}}, \frac{{{\alpha _T}}}{{{v_{\rm{T}}}}},\frac{{{\alpha _T}^2}}{{{v_{\rm{T}}}}},\frac{{{\alpha _T}^3}}{{{v_{\rm{T}}}}},\frac{{{\alpha _T}^4}}{{{v_{\rm{T}}}}},\frac{{{\alpha _T}^5}}{{{v_{\rm{T}}}}},{\alpha _T}\dot \psi ,\\
 \quad \quad {\alpha _{\rm{T}}}^2\dot \psi,{\alpha _{\rm{T}}}^3\dot \psi,{\alpha _{\rm{T}}}^4\dot \psi ,{\alpha _{\rm{T}}}^5\dot \psi,\frac{1}{{{v_{\rm{T}}}}}\dot \psi ,\frac{{{\alpha _T}}}{{{v_{\rm{T}}}}}\dot \psi ,\frac{{{\alpha _T}^2}}{{{v_{\rm{T}}}}}\dot \psi,\frac{{{\alpha _T}^3}}{{{v_{\rm{T}}}}}\dot \psi,\frac{{{\alpha _T}^4}}{{{v_{\rm{T}}}}}\dot \psi,\frac{{{\alpha _T}^5}}{{{v_{\rm{T}}}}}\dot \psi \} 
\end{array}}
\label{eq:60e2}
\end{align}

\begin{figure}[!t]
	\centering{\includegraphics [width=3.5in] {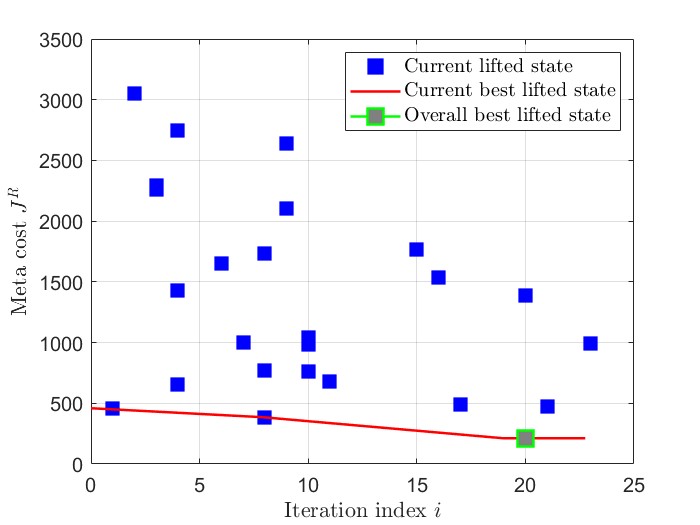}}
	\caption{Meta cost $J^R$ vs iteration $i$ of Algorithm 1. At each iteration $i$, the cost of the current set of observable functions and of the current best one are depicted. The green square shows the overall best lifted state $\xi _{\theta}= [\psi, {\alpha _T},\dot \psi ,{\alpha _T}^2,{\alpha _{\rm{T}}}^3,\frac{{{\alpha _T}}}{{{v_{\rm{T}}}}},\frac{{{\alpha _T}^2}}{{{v_{\rm{T}}}}},\frac{{{\alpha _T}^3}}{{{v_{\rm{T}}}}},{\alpha _T}\dot \psi ,{\alpha _{\rm{T}}}^2\dot \psi , \allowbreak \frac{1}{{{v_{\rm{T}}}}}\dot \psi ,  \frac{{{\alpha _T}}}{{{v_{\rm{T}}}}}\dot \psi ,\frac{{{\alpha _T}^2}}{{{v_{\rm{T}}}}}\dot \psi ]^{\top}$ which is found at iteration $20$.}  \label{FigBysE21}
\end{figure}

{The meta cost $J^R$ vs iteration $i$ of Algorithm 1 is illustrated in Fig.~\ref{FigBysE21}. Note that for the lifted state $\xi _{\theta}= [\psi, {\alpha _T},\dot \psi ,{\alpha _T}^2,{\alpha _{\rm{T}}}^3,\frac{{{\alpha _T}}}{{{v_{\rm{T}}}}},\frac{{{\alpha _T}^2}}{{{v_{\rm{T}}}}},\frac{{{\alpha _T}^3}}{{{v_{\rm{T}}}}},{\alpha _T}\dot \psi ,{\alpha _{\rm{T}}}^2\dot \psi , \allowbreak \frac{1}{{{v_{\rm{T}}}}}\dot \psi ,  \frac{{{\alpha _T}}}{{{v_{\rm{T}}}}}\dot \psi ,\frac{{{\alpha _T}^2}}{{{v_{\rm{T}}}}}\dot \psi ]^{\top}$, one has 
\begin{align}
	\Scale[0.95]{\begin{array}{l}
			{{\dot \xi }_\theta } = {{\hat A}^*}{\xi _\theta } + \hat B_1^*{\xi _\theta }\delta  + \hat B_2^*{\xi _\theta }sin(\delta ) + \hat B_3^*{\xi _\theta }\delta sin(\delta ) + \hat B_4^*{\xi _\theta }{\delta ^2} + \hat B_5^*{\xi _\theta }{\delta ^3} + {{\hat {\cal P}}_\theta }\left( {t,{\xi _\theta }(t)} \right)\\
			\,\,\,\,\,\,\,\,\, = {{\hat A}^*}{\xi _\theta } + \underbrace {\left[ {\begin{array}{*{20}{c}}
						{{I_{{n_{{\xi _\theta }}}}}}&{{I_{{n_{{\xi _\theta }}}}}}&{{I_{{n_{{\xi _\theta }}}}}}&{\begin{array}{*{20}{c}}
								{{I_{{n_{{\xi _\theta }}}}}}&{{I_{{n_{{\xi _\theta }}}}}}
						\end{array}}
				\end{array}} \right]diag(\hat B_1^*,\hat B_2^*,\hat B_3^*,\hat B_4^*,\hat B_5^*)}_{\hat B_\theta ^*}\underbrace {({I_5} \otimes {\xi _\theta })\left[ {\begin{array}{*{20}{c}}
						\delta \\
						{\sin \delta }\\
						{\delta \sin \delta }\\
						{\begin{array}{*{20}{c}}
								{{\delta ^2}}\\
								{{\delta ^3}}
						\end{array}}
				\end{array}} \right]}_{\Psi ({\cal U}(\delta (t)))} + {{\hat {\cal P}}_\theta }\left( {t,{\xi _\theta }(t)} \right)
		\end{array}}
	\label{eq:62e2}
\end{align}}
{\begin{align}
\begin{array}{l}
	{{\dot v}_{\rm{T}}} = {\left[ {0,0,0,\frac{{2K}}{{{m_T}}},0,0,0,0,\frac{{K(a - b)}}{{{m_T}{v_{{\rm{T}},0}}}},0,0,0,0} \right]^ \top }{\xi _\theta }\\
	\,\,\,\,\,\, + \left[ {\begin{array}{*{20}{c}}
			{\begin{array}{*{20}{c}}
					{{{\bar {\cal G}}_{11}}}\\
					0\\
					0\\
					{\begin{array}{*{20}{c}}
							0\\
							0
					\end{array}}
			\end{array}}&{\begin{array}{*{20}{c}}
					0\\
					{{{\bar {\cal G}}_{12}}}\\
					0\\
					{\begin{array}{*{20}{c}}
							0\\
							0
					\end{array}}
			\end{array}}&{\begin{array}{*{20}{c}}
					0\\
					0\\
					{{{\bar {\cal G}}_{13}}}\\
					{\begin{array}{*{20}{c}}
							0\\
							0
					\end{array}}
			\end{array}}&{\begin{array}{*{20}{c}}
					{\begin{array}{*{20}{c}}
							0\\
							0\\
							0\\
							{\begin{array}{*{20}{c}}
									{{{\bar {\cal G}}_{14}}}\\
									0
							\end{array}}
					\end{array}}&{\begin{array}{*{20}{c}}
							0\\
							0\\
							0\\
							{\begin{array}{*{20}{c}}
									0\\
									{{{\bar {\cal G}}_{15}}}
							\end{array}}
					\end{array}}
			\end{array}}
	\end{array}} \right]({I_5} \otimes {\xi _\theta })\left[ {\begin{array}{*{20}{c}}
			\delta \\
			{\sin \delta }\\
			{\delta \sin \delta }\\
			{\begin{array}{*{20}{c}}
					{{\delta ^2}}\\
					{{\delta ^3}}
			\end{array}}
	\end{array}} \right]
\end{array} \label{eq:62e3}
\end{align}}
{where ${{\bar {\cal G}}_{11}} = \left[ {0,\frac{{ - K}}{{{m_T}}},0,0,0,0,0,0,0,0,0,0,0} \right]$, ${{\bar {\cal G}}_{12}} = \left[ {0,\frac{{ - K}}{{{m_T}}},\frac{{ - Ka}}{{{m_T}{v_{{\rm{T}},0}}}},0,\frac{K}{{2{m_T}}},0,0,0,0,\frac{{Ka}}{{2{m_T}{v_{{\rm{T}},0}}}},0,0,0} \right]$, ${{\bar {\cal G}}_{13}} = \left[ {0,\frac{K}{{{m_T}}},0,\frac{{ - K}}{{2{m_T}}},0,0,0,0,0,0,0,0,0} \right]$, ${{\bar {\cal G}}_{14}} = \left[ {0,0,0,\frac{{ - K}}{{2{m_T}}},0,0,0,0,\frac{{ - Ka}}{{2{m_T}{v_{{\rm{T}},0}}}},0,0,0,0} \right]$, and ${{\bar {\cal G}}_{15}} = \left[ {0,\frac{K}{{2{m_T}}},0,0,0,0,0,0,0,0,0,0,0} \right]$.}


\begin{figure}[!t]
\centering{\includegraphics [width=3in] {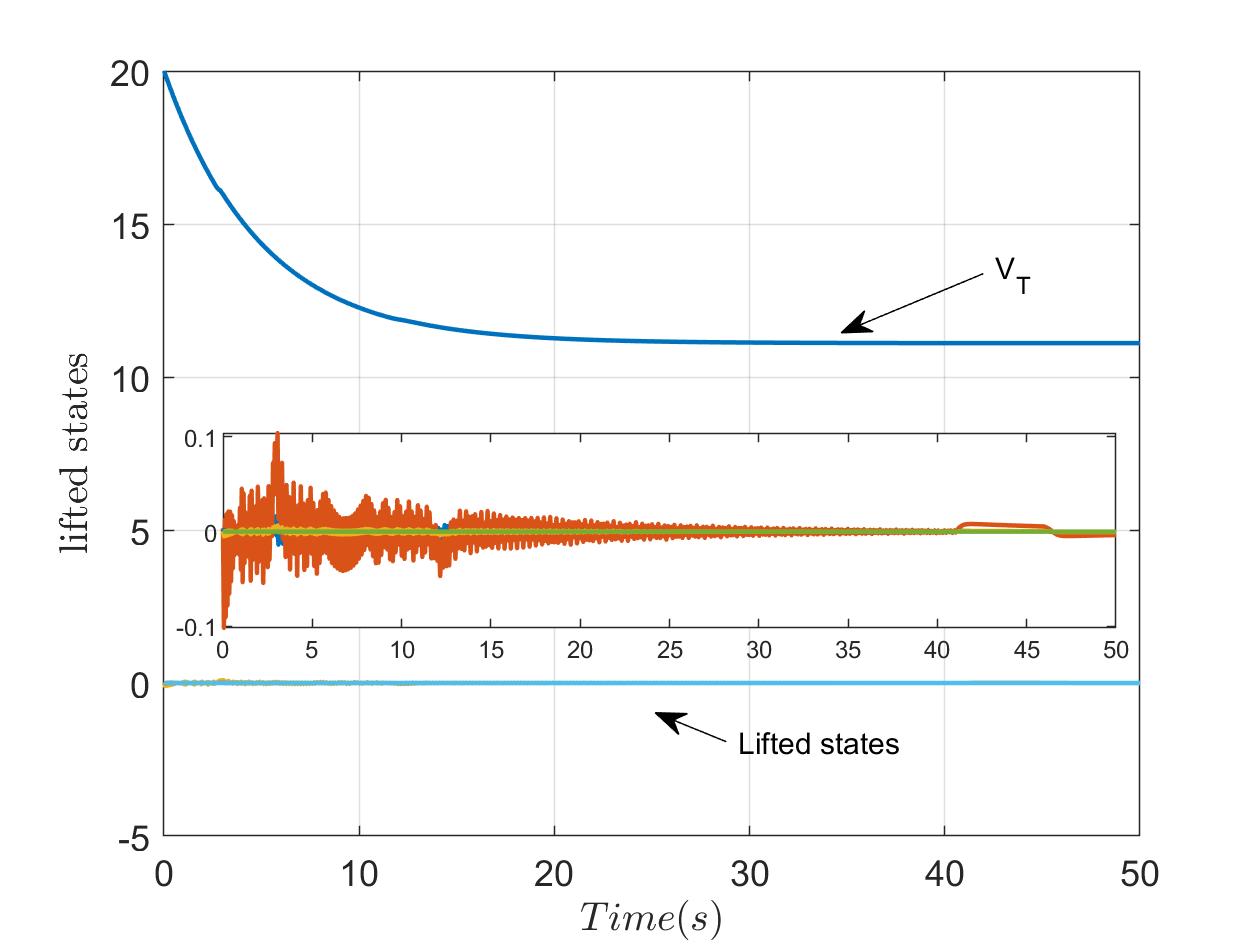}}
\caption{Evolution of the lifted state $\xi_{\theta}$ and $v_T$ trajectories.}
\label{fig:3e2}
\end{figure}

\begin{figure}[!t]
\centering{\includegraphics [width=3in] {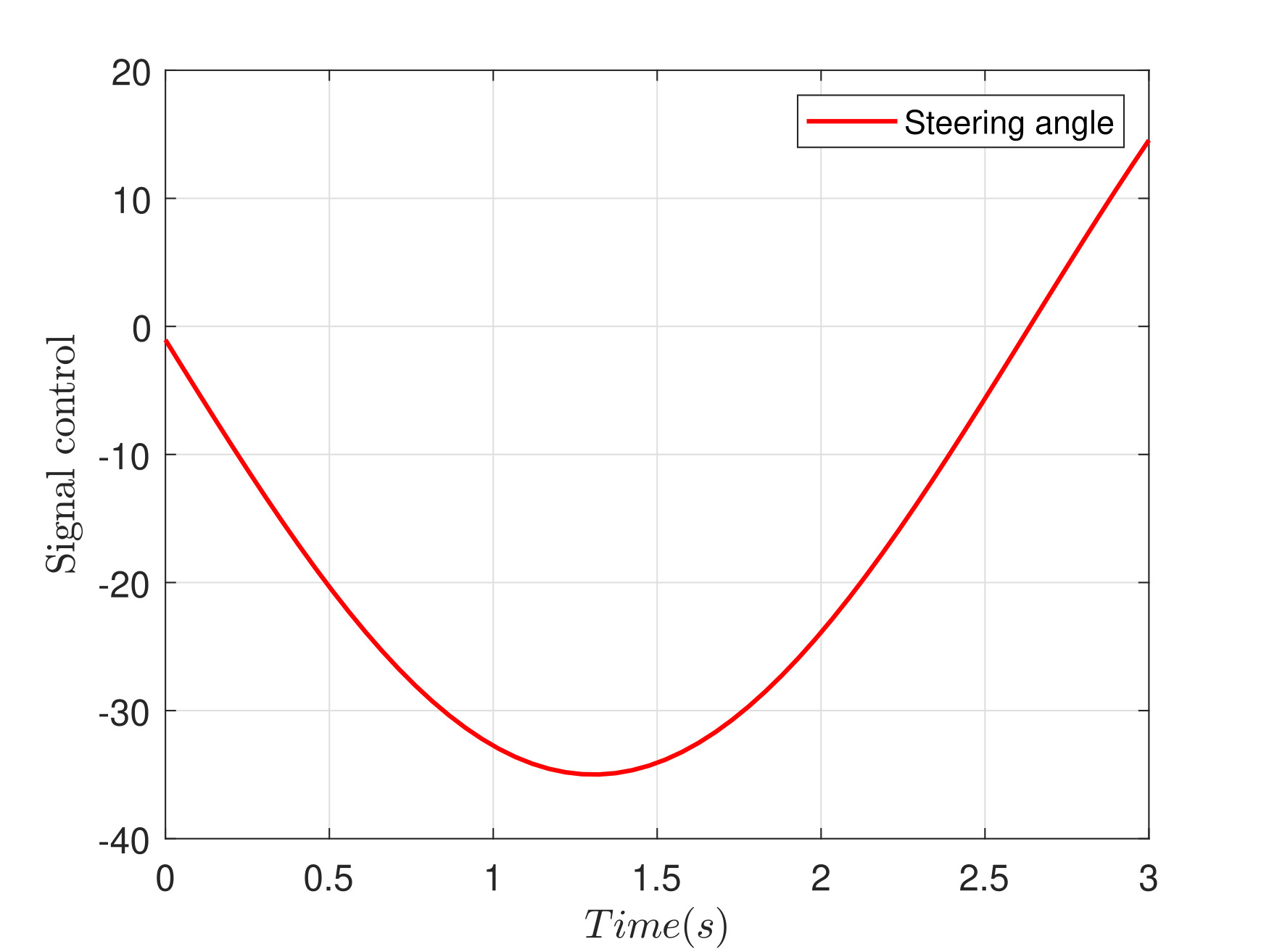}}
\caption{Control input $\delta = -35Sin(1.2t)$ Degree.}
\label{fig:6e2}
\end{figure}

\begin{figure}[!t]
\centering{\includegraphics [width=3in] {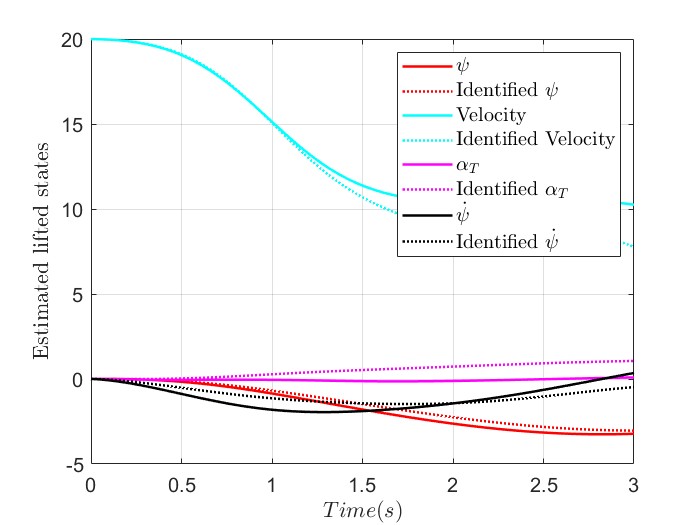}}
\caption{States trajectories of the nonlinear vehicle system and the identified model under control input of $\delta = -35Sin(1.2t)$ Degree.}
\label{fig:7e2}
\end{figure}

\begin{figure}[!t]
\centering{\includegraphics [width=2.8in] {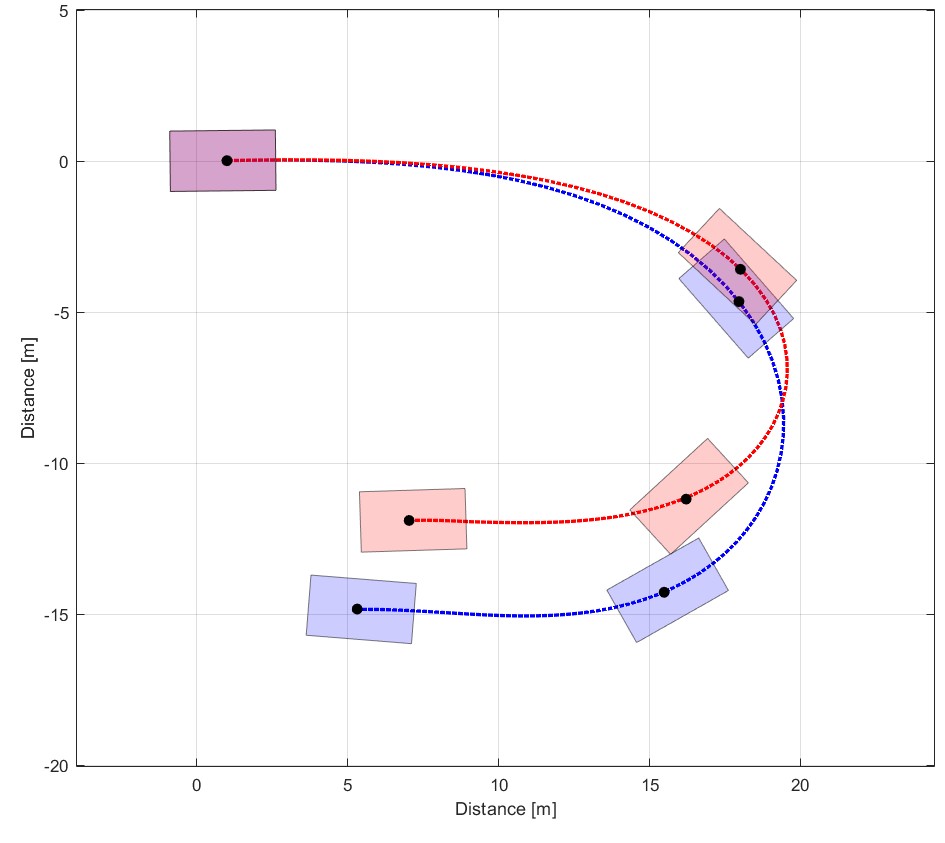}}
\caption{$X$ and $Y$ position trajectories of the nonlinear vehicle system (Light-Blue square) and the identified model (Light-Coral square) under control input of $\delta = -35Sin(1.2t)$ Degree.}
\label{fig:8e2}
\end{figure}

To satisfy Condition 1, a rich enough input signal $\delta$ is injected into the nonlinear vehicle system \eqref{eq:9e2} for excitation of the lifted states $\xi_ \theta$. 
The evolution of the lifted states trajectories under the rich enough input signal $\delta$ is shown in Fig.~\ref{fig:3e2}. The proposed update law \eqref{eq:29}-\eqref{eq:31} is used to find the approximated matrices $\hat A^{*}_{\theta}$, and $\hat B^{*}_{\theta}$. After the convergence of $\hat A_\theta ^*$ and $\hat B_\theta ^*$, the control input $\delta$, given in Fig.~\ref{fig:6e2}, is employed for the identified linear representation system  \eqref{eq:62e2} and the nonlinear vehicle system \eqref{eq:9e2} to compare the evolution of the states trajectories of the identified model and the nonlinear vehicle system \eqref{eq:9e2}. Using the equations of motion  \citep{andrzejewski2006nonlinear,de2016vehicle}
\begin{align}
\Scale[1]{ \begin{array}{l}
{\rm{\dot X  =  }}{{\rm{v}}_T}{\rm{cos(}}{\alpha _T}{\rm{  +  }}\psi {\rm{)}}, \\
{\rm{\dot Y  =  }}{{\rm{v}}_T}{\rm{sin(}}{\alpha _T}{\rm{  +  }}\psi {\rm{)}}.\\
\end{array}}
\label{eq:70e2}
\end{align}
 Fig. \ref{fig:8e2}, illustrates the comparison of $X-Y$ positions of the nonlinear vehicle system \eqref{eq:9e2} and the identified model \eqref{eq:62e2}.

\begin{figure}[!t]
\centering{\includegraphics [width=3in] {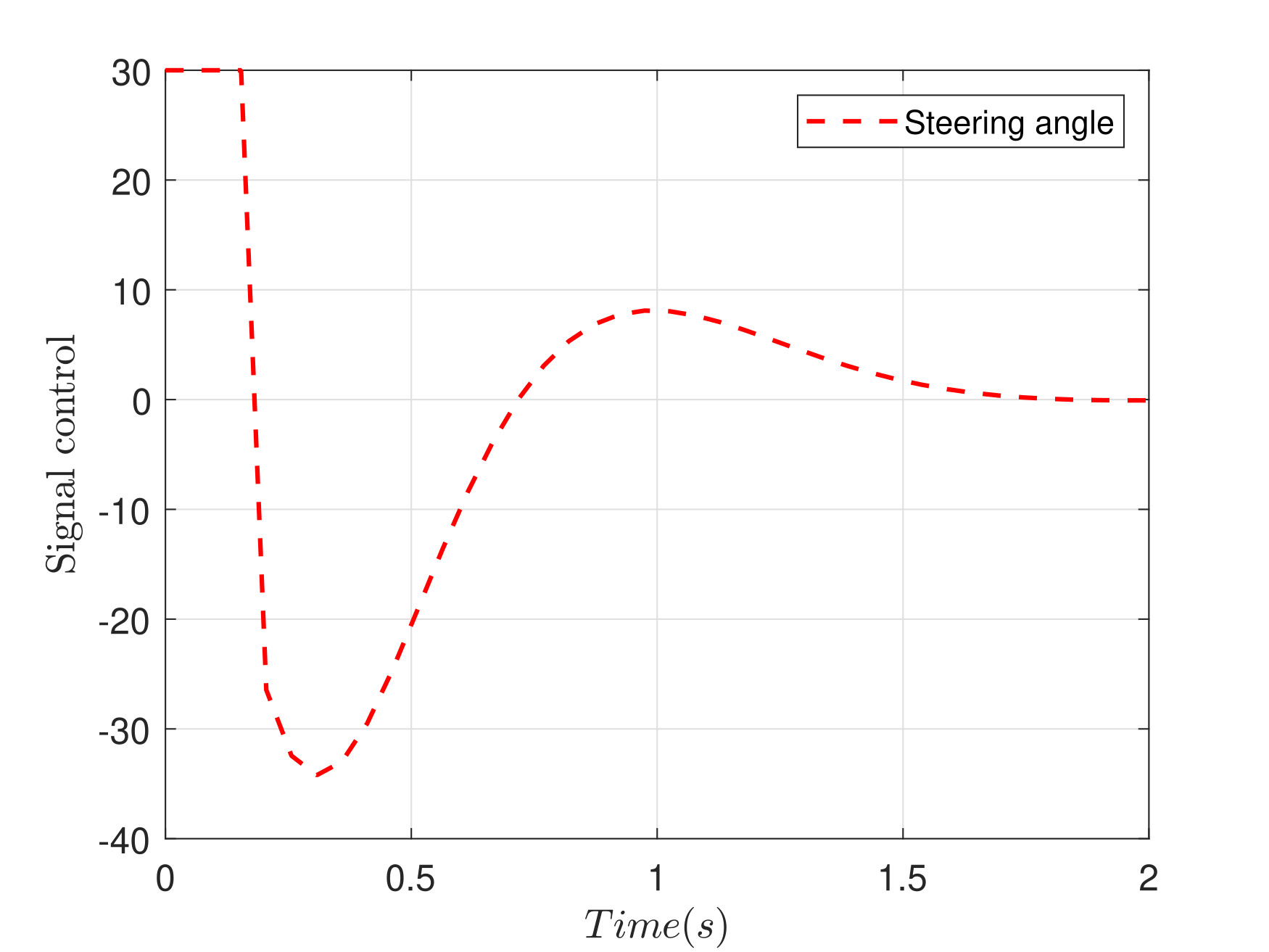}}
\caption{Control input $\delta$.}
\label{fig:11e2}
\end{figure}

\begin{figure}[!t]
\centering{\includegraphics [width=3in] {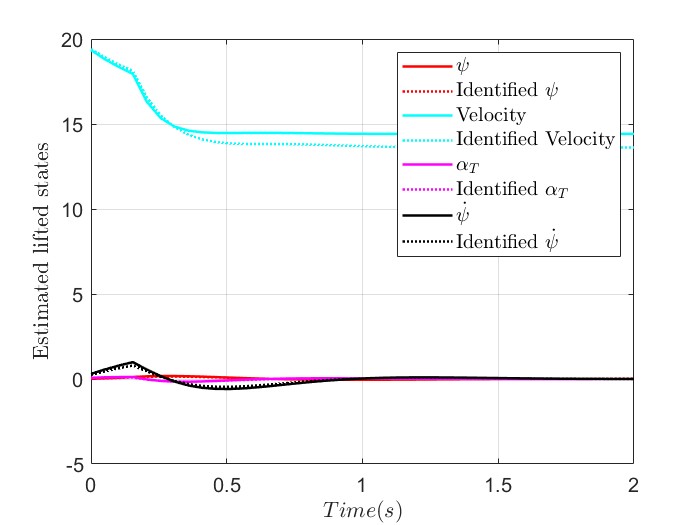}}
\caption{State trajectories of identified  { (dashed line) } and real { (solid line) } nonlinear system with $\delta$.}
\label{fig:12e2}
\end{figure}

\begin{figure}[!t]
\centering{\includegraphics [width=3.3in] {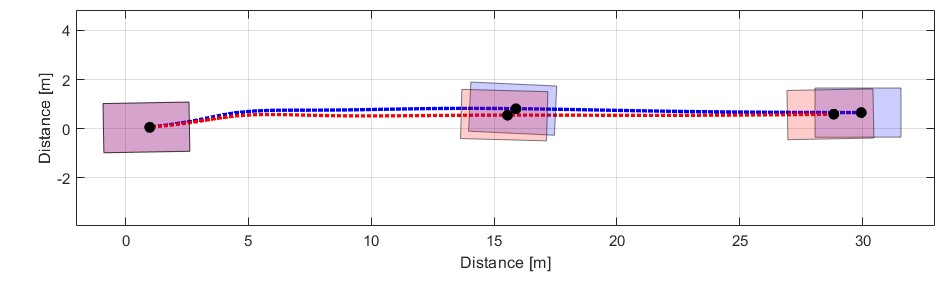}}
\caption{$X$ and $Y$ position trajectories of the nonlinear vehicle system (Light-Blue square) and the identified model (Light-Coral square) with control input of $\delta $.}
\label{fig:13e2}
\end{figure}

Now, to  test  the  learned model, we assumed that  a  disturbance  as  a  single steering pulse  with the magnitude of $\delta = 30^\circ$ is simulated at $t=0 - 0.2$ sec. The control input $\delta$ as depicted in Fig.~\ref{fig:11e2}  is employed for the nonlinear vehicle system \eqref{eq:9e2} to stabilize the states $\{\psi,\alpha_T,\dot \psi\}$. For the comparison, the feedback gain $K$  is also applied to the identified model.
 Fig.~\ref{fig:12e2} depicts the evolution of state trajectories of identified  { (dashed line) } and real { (solid line) } nonlinear system, which shows that the both state trajectories.  Fig. \ref{fig:13e2}, illustrates the comparison of $X-Y$ positions of the nonlinear vehicle system \eqref{eq:9} and the identified model \eqref{eq:62} under the control input given in Fig.~\ref{fig:13e2}.

\section{Conclusion}
A novel data-driven learning algorithm is presented to learn a linear representation of nonlinear system dynamics using Koopman operator theory. To jointly learn the set of observables (structure) and Koopman parameters, a bilevel learning mechanism with two layers of learning is developed. The lower layer learner leverages unified batch-online learning-based finite-time Koopman identifier which uses discontinuous gradient update laws to minimize the instantaneous Koopman operator's identification errors as well as the identification errors for a batch of past samples collected in a history stack.
It is guaranteed that the lower layer identifier will converge in a finite time under easy-to-verify conditions on a batch of samples.
 A higher layer employs a discrete Bayesian optimization algorithm to find a set of observables with minimum approximation errors and minimum cardinality. Finally, the effectiveness of the proposed framework was verified on a simulation example.



\acks{This work was supported by Ford Motor Company-Michigan State University Alliance. }



\bibliography{REFERENCE.bib}

\begin{thebibliography}{66}
\providecommand{\natexlab}[1]{#1}
\providecommand{\url}[1]{\texttt{#1}}
\expandafter\ifx\csname urlstyle\endcsname\relax
  \providecommand{\doi}[1]{doi: #1}\else
  \providecommand{\doi}{doi: \begingroup \urlstyle{rm}\Url}\fi

\bibitem[Andrzejewski and Awrejcewicz(2006)]{andrzejewski2006nonlinear}
Ryszard Andrzejewski and Jan Awrejcewicz.
\newblock \emph{Nonlinear dynamics of a wheeled vehicle}, volume~10.
\newblock Springer Science \& Business Media, 2006.

\bibitem[Ayoobi et~al.(2021)Ayoobi, Cao, Verbrugge, and
  Verheij]{ayoobi2021argumentation}
Hamed Ayoobi, Ming Cao, Rineke Verbrugge, and Bart Verheij.
\newblock Argumentation-based online incremental learning.
\newblock \emph{IEEE Transactions on Automation Science and Engineering}, pages
  1--15, 2021.

\bibitem[Bacciotti and Ceragioli(1999)]{bacciotti1999stability}
Andrea Bacciotti and Francesca Ceragioli.
\newblock Stability and stabilization of discontinuous systems and nonsmooth
  lyapunov functions.
\newblock \emph{ESAIM: Control, Optimisation and Calculus of Variations},
  4:\penalty0 361--376, 1999.

\bibitem[Bakker et~al.(2019)Bakker, Rosenthal, and Nowak]{bakker2019koopman}
Craig Bakker, Steven Rosenthal, and Kathleen~E Nowak.
\newblock Koopman representations of dynamic systems with control.
\newblock \emph{arXiv preprint arXiv:1908.02233}, 2019.

\bibitem[Blockeel et~al.(2013)Blockeel, Kersting, Nijssen, and
  {\v{Z}}elezn{\`y}]{blockeel2013machine}
Hendrik Blockeel, Kristian Kersting, Siegfried Nijssen, and Filip
  {\v{Z}}elezn{\`y}.
\newblock \emph{Machine Learning and Knowledge Discovery in Databases: European
  Conference, ECML PKDD 2013, Prague, Czech Republic, September 23-27, 2013,
  Proceedings, Part I}, volume 8188.
\newblock Springer, 2013.

\bibitem[Brochu et~al.(2010)Brochu, Cora, and De~Freitas]{brochu2010tutorial}
Eric Brochu, Vlad~M Cora, and Nando De~Freitas.
\newblock A tutorial on bayesian optimization of expensive cost functions, with
  application to active user modeling and hierarchical reinforcement learning.
\newblock \emph{arXiv preprint arXiv:1012.2599}, 2010.

\bibitem[Brunton et~al.(2016{\natexlab{a}})Brunton, Brunton, Proctor, and
  Kutz]{Brunton2016KoopmanIS}
Steven~L. Brunton, B.~Brunton, Joshua~L. Proctor, and J.~Nathan Kutz.
\newblock Koopman invariant subspaces and finite linear representations of
  nonlinear dynamical systems for control.
\newblock \emph{PLoS ONE}, 11, 2016{\natexlab{a}}.

\bibitem[Brunton et~al.(2016{\natexlab{b}})Brunton, Brunton, Proctor, and
  Kutz]{brunton2016koopman}
Steven~L Brunton, Bingni~W Brunton, Joshua~L Proctor, and J~Nathan Kutz.
\newblock Koopman invariant subspaces and finite linear representations of
  nonlinear dynamical systems for control.
\newblock \emph{PloS one}, 11\penalty0 (2):\penalty0 e0150171,
  2016{\natexlab{b}}.

\bibitem[Budi{\v{s}}i{\'c} et~al.(2012)Budi{\v{s}}i{\'c}, Mohr, and
  Mezi{\'c}]{budivsic2012applied}
Marko Budi{\v{s}}i{\'c}, Ryan Mohr, and Igor Mezi{\'c}.
\newblock Applied koopmanism.
\newblock \emph{Chaos: An Interdisciplinary Journal of Nonlinear Science},
  22\penalty0 (4):\penalty0 047510, 2012.

\bibitem[Cao et~al.(2017)Cao, Xiao, Huang, and Zhou]{cao2017robust}
Zhengcai Cao, Qing Xiao, Ran Huang, and Mengchu Zhou.
\newblock Robust neuro-optimal control of underactuated snake robots with
  experience replay.
\newblock \emph{IEEE transactions on neural networks and learning systems},
  29\penalty0 (1):\penalty0 208--217, 2017.

\bibitem[Chen et~al.(2021)Chen, Deng, and Yang]{chen2021practical}
Guopei Chen, Feiqi Deng, and Ying Yang.
\newblock Practical finite-time stability of switched nonlinear time-varying
  systems based on initial state-dependent dwell time methods.
\newblock \emph{Nonlinear Analysis: Hybrid systems}, 41:\penalty0 101031, 2021.

\bibitem[Chowdhary and Johnson(2010)]{chowdhary2010concurrent}
Girish Chowdhary and Eric Johnson.
\newblock Concurrent learning for convergence in adaptive control without
  persistency of excitation.
\newblock In \emph{49th IEEE Conference on Decision and Control (CDC)}, pages
  3674--3679. IEEE, 2010.

\bibitem[Chowdhary and Johnson(2011)]{chowdhary2011singular}
Girish Chowdhary and Eric Johnson.
\newblock A singular value maximizing data recording algorithm for concurrent
  learning.
\newblock In \emph{Proceedings of the 2011 American Control Conference}, pages
  3547--3552. IEEE, 2011.

\bibitem[Chowdhary et~al.(2013)Chowdhary, Yucelen, M{\"u}hlegg, and
  Johnson]{chowdhary2013concurrent}
Girish Chowdhary, Tansel Yucelen, Maximillian M{\"u}hlegg, and Eric~N Johnson.
\newblock Concurrent learning adaptive control of linear systems with
  exponentially convergent bounds.
\newblock \emph{International Journal of Adaptive Control and Signal
  Processing}, 27\penalty0 (4):\penalty0 280--301, 2013.

\bibitem[Cort{\'e}s(2008)]{Corts2008DiscontinuousDS}
Jorge Cort{\'e}s.
\newblock Discontinuous dynamical systems.
\newblock \emph{IEEE Control Systems}, 28, 2008.

\bibitem[de~Souza~Mendes et~al.(2016)de~Souza~Mendes, Meneghetti, Ackermann,
  and de~Toledo~Fleury]{de2016vehicle}
Andr{\'e} de~Souza~Mendes, Douglas De~Rizzo Meneghetti, Marko Ackermann, and
  Agenor de~Toledo~Fleury.
\newblock Vehicle dynamics-lateral: Open source simulation package for matlab.
\newblock Technical report, SAE Technical Paper, 2016.

\bibitem[Drma{\v{c}} et~al.(2021)Drma{\v{c}}, Mezi{\'c}, and
  Mohr]{drmavc2021identification}
Zlatko Drma{\v{c}}, Igor Mezi{\'c}, and Ryan Mohr.
\newblock Identification of nonlinear systems using the infinitesimal generator
  of the koopman semigroup—a numerical implementation of the
  mauroy--goncalves method.
\newblock \emph{Mathematics}, 9\penalty0 (17):\penalty0 2075, 2021.

\bibitem[Filippov and Arscott(1988)]{filippov1988differential}
A.~F. Filippov and F.~M. Arscott.
\newblock \emph{Differential Equations with Discontinuous Righthand Sides:
  Control Systems (Mathematics and its Applications, 18)}.
\newblock Dordrecht, Netherlands: Kluwer Academic Publishers Group., 1988.

\bibitem[Filippov(2013)]{filippov2013differential}
Aleksei~Fedorovich Filippov.
\newblock \emph{Differential equations with discontinuous righthand sides:
  control systems}, volume~18.
\newblock Springer Science \& Business Media, 2013.

\bibitem[Han et~al.(2020)Han, Hao, and Vaidya]{han2020deep}
Yiqiang Han, Wenjian Hao, and Umesh Vaidya.
\newblock Deep learning of koopman representation for control.
\newblock In \emph{2020 59th IEEE Conference on Decision and Control (CDC)},
  pages 1890--1895. IEEE, 2020.

\bibitem[He and Song(2017)]{he2017finite}
Shuping He and Jun Song.
\newblock Finite-time sliding mode control design for a class of uncertain
  conic nonlinear systems.
\newblock \emph{IEEE/CAA Journal of Automatica Sinica}, 4\penalty0
  (4):\penalty0 809--816, 2017.

\bibitem[Huang et~al.(2020)Huang, Ma, and Vaidya]{huang2020data}
Bowen Huang, Xu~Ma, and Umesh Vaidya.
\newblock Data-driven nonlinear stabilization using koopman operator.
\newblock In \emph{The Koopman Operator in Systems and Control}, pages
  313--334. Springer, 2020.

\bibitem[Jha et~al.(2019)Jha, Roy, and Bhasin]{jha2019initial}
Sumit~Kumar Jha, Sayan~Basu Roy, and Shubhendu Bhasin.
\newblock Initial excitation-based iterative algorithm for approximate optimal
  control of completely unknown lti systems.
\newblock \emph{IEEE Transactions on Automatic Control}, 64\penalty0
  (12):\penalty0 5230--5237, 2019.

\bibitem[Jiang et~al.(2019)Jiang, Huang, and Ding]{jiang2019path}
Lan Jiang, Hongyun Huang, and Zuohua Ding.
\newblock Path planning for intelligent robots based on deep q-learning with
  experience replay and heuristic knowledge.
\newblock \emph{IEEE/CAA Journal of Automatica Sinica}, 7\penalty0
  (4):\penalty0 1179--1189, 2019.

\bibitem[Kaiser et~al.(2021)Kaiser, Kutz, and Brunton]{kaiser2021data}
Eurika Kaiser, J~Nathan Kutz, and Steven Brunton.
\newblock Data-driven discovery of koopman eigenfunctions for control.
\newblock \emph{Machine Learning: Science and Technology}, 2021.

\bibitem[Kamalapurkar et~al.(2017)Kamalapurkar, Reish, Chowdhary, and
  Dixon]{kamalapurkar2017concurrent}
Rushikesh Kamalapurkar, Benjamin Reish, Girish Chowdhary, and Warren~E Dixon.
\newblock Concurrent learning for parameter estimation using dynamic
  state-derivative estimators.
\newblock \emph{IEEE Transactions on Automatic Control}, 62\penalty0
  (7):\penalty0 3594--3601, 2017.

\bibitem[Korda and Mezi{\'c}(2018)]{korda2018linear}
Milan Korda and Igor Mezi{\'c}.
\newblock Linear predictors for nonlinear dynamical systems: Koopman operator
  meets model predictive control.
\newblock \emph{Automatica}, 93:\penalty0 149--160, 2018.

\bibitem[Kutz et~al.(2016)Kutz, Proctor, and Brunton]{kutz2016koopman}
J~Nathan Kutz, Joshua~L Proctor, and Steven~L Brunton.
\newblock Koopman theory for partial differential equations.
\newblock \emph{arXiv preprint arXiv:1607.07076}, 2016.

\bibitem[Lehrer et~al.(2010)Lehrer, Adetola, and Guay]{lehrer2010parameter}
Devon Lehrer, Veronica Adetola, and Martin Guay.
\newblock Parameter identification methods for non-linear discrete-time
  systems.
\newblock In \emph{Proceedings of the 2010 American Control Conference}, pages
  2170--2175. IEEE, 2010.

\bibitem[Li et~al.(2017)Li, Dietrich, Bollt, and Kevrekidis]{li2017extended}
Qianxiao Li, Felix Dietrich, Erik~M Bollt, and Ioannis~G Kevrekidis.
\newblock Extended dynamic mode decomposition with dictionary learning: A
  data-driven adaptive spectral decomposition of the koopman operator.
\newblock \emph{Chaos: An Interdisciplinary Journal of Nonlinear Science},
  27\penalty0 (10):\penalty0 103111, 2017.

\bibitem[Liu et~al.(2014)Liu, Zhou, Zhu, Fu, and Fu]{liu2014experience}
Quan Liu, Xin Zhou, Fei Zhu, Qiming Fu, and Yuchen Fu.
\newblock Experience replay for least-squares policy iteration.
\newblock \emph{IEEE/CAA Journal of Automatica Sinica}, 1\penalty0
  (3):\penalty0 274--281, 2014.

\bibitem[Liu et~al.(2021)Liu, Liu, Jing, and Zhang]{liu2021semi}
Yang Liu, Xiaoping Liu, Yuanwei Jing, and Ziye Zhang.
\newblock Semi-globally practical finite-time stability for uncertain nonlinear
  systems based on dynamic surface control.
\newblock \emph{International Journal of Control}, 94\penalty0 (2):\penalty0
  476--485, 2021.

\bibitem[Lu et~al.(2016)Lu, Liu, and Chen]{lu2016note}
Wenlian Lu, Xiwei Liu, and Tianping Chen.
\newblock A note on finite-time and fixed-time stability.
\newblock \emph{Neural Networks}, 81:\penalty0 11--15, 2016.

\bibitem[Luong et~al.(2019)Luong, Gupta, Nguyen, Rana, and
  Venkatesh]{luong2019bayesian}
Phuc Luong, Sunil Gupta, Dang Nguyen, Santu Rana, and Svetha Venkatesh.
\newblock Bayesian optimization with discrete variables.
\newblock In \emph{Australasian Joint Conference on Artificial Intelligence},
  pages 473--484. Springer, 2019.

\bibitem[Maghenem and Sanfelice(2018)]{maghenem2018barrier}
Mohamed Maghenem and Ricardo~G Sanfelice.
\newblock Barrier function certificates for forward invariance in hybrid
  inclusions.
\newblock In \emph{2018 IEEE Conference on Decision and Control (CDC)}, pages
  759--764. IEEE, 2018.

\bibitem[Mauroy and Goncalves(2019)]{mauroy2019koopman}
Alexandre Mauroy and Jorge Goncalves.
\newblock Koopman-based lifting techniques for nonlinear systems
  identification.
\newblock \emph{IEEE Transactions on Automatic Control}, 65\penalty0
  (6):\penalty0 2550--2565, 2019.

\bibitem[Modares et~al.(2013)Modares, Lewis, and
  Naghibi-Sistani]{modares2013adaptive}
Hamidreza Modares, Frank~L Lewis, and Mohammad-Bagher Naghibi-Sistani.
\newblock Adaptive optimal control of unknown constrained-input systems using
  policy iteration and neural networks.
\newblock \emph{IEEE transactions on neural networks and learning systems},
  24\penalty0 (10):\penalty0 1513--1525, 2013.

\bibitem[Netto and Mili(2018)]{netto2018robust}
Marcos Netto and Lamine Mili.
\newblock A robust data-driven koopman kalman filter for power systems dynamic
  state estimation.
\newblock \emph{IEEE Transactions on Power Systems}, 33\penalty0 (6):\penalty0
  7228--7237, 2018.

\bibitem[Paden and Sastry(1987)]{paden1987calculus}
Brad Paden and Shankar Sastry.
\newblock A calculus for computing filippov's differential inclusion with
  application to the variable structure control of robot manipulators.
\newblock \emph{IEEE transactions on circuits and systems}, 34\penalty0
  (1):\penalty0 73--82, 1987.

\bibitem[Parikh et~al.(2019)Parikh, Kamalapurkar, and
  Dixon]{parikh2019integral}
Anup Parikh, Rushikesh Kamalapurkar, and Warren~E Dixon.
\newblock Integral concurrent learning: Adaptive control with parameter
  convergence using finite excitation.
\newblock \emph{International Journal of Adaptive Control and Signal
  Processing}, 33\penalty0 (12):\penalty0 1775--1787, 2019.

\bibitem[Proctor et~al.(2018)Proctor, Brunton, and
  Kutz]{proctor2018generalizing}
Joshua~L Proctor, Steven~L Brunton, and J~Nathan Kutz.
\newblock Generalizing koopman theory to allow for inputs and control.
\newblock \emph{SIAM Journal on Applied Dynamical Systems}, 17\penalty0
  (1):\penalty0 909--930, 2018.

\bibitem[Romero and Benosman(2020{\natexlab{a}})]{Romero2020RobustTC}
Orlando Romero and Mouhacine Benosman.
\newblock Robust time-varying continuous-time optimization with pre-defined
  finite-time stability.
\newblock \emph{IFAC-PapersOnLine}, 53:\penalty0 6743--6748,
  2020{\natexlab{a}}.

\bibitem[Romero and Benosman(2020{\natexlab{b}})]{romero2020finite}
Orlando Romero and Mouhacine Benosman.
\newblock Finite-time convergence in continuous-time optimization.
\newblock In \emph{International Conference on Machine Learning}, pages
  8200--8209. PMLR, 2020{\natexlab{b}}.

\bibitem[Romero and Benosman(2021)]{Romero2021TimevaryingCO}
Orlando Romero and Mouhacine Benosman.
\newblock Time-varying continuous-time optimisation with pre-defined
  finite-time stability.
\newblock \emph{International Journal of Control}, 94:\penalty0 3237 -- 3254,
  2021.

\bibitem[Shahriari et~al.(2015)Shahriari, Swersky, Wang, Adams, and
  De~Freitas]{shahriari2015taking}
Bobak Shahriari, Kevin Swersky, Ziyu Wang, Ryan~P Adams, and Nando De~Freitas.
\newblock Taking the human out of the loop: A review of bayesian optimization.
\newblock \emph{Proceedings of the IEEE}, 104\penalty0 (1):\penalty0 148--175,
  2015.

\bibitem[Song and Wei(2019)]{song2019neural}
Ruizhuo Song and Qinglai Wei.
\newblock Neural-network-based approach for finite-time optimal control.
\newblock In \emph{Adaptive Dynamic Programming: Single and Multiple
  Controllers}, pages 7--23. Springer, 2019.

\bibitem[Tao(2003)]{tao2003adaptive}
Gang Tao.
\newblock \emph{Adaptive control design and analysis}, volume~37.
\newblock John Wiley \& Sons, 2003.

\bibitem[Tatari et~al.(2017)Tatari, Naghibi-Sistani, and
  Vamvoudakis]{tatari2017distributed}
Farzaneh Tatari, Mohammad-Bagher Naghibi-Sistani, and Kyriakos~G Vamvoudakis.
\newblock Distributed optimal synchronization control of linear networked
  systems under unknown dynamics.
\newblock In \emph{2017 American Control Conference (ACC)}, pages 668--673.
  IEEE, 2017.

\bibitem[Tatari et~al.(2018)Tatari, Vamvoudakis, and
  Mazouchi]{tatari2018optimal}
Farzaneh Tatari, Kyriakos~G Vamvoudakis, and Majid Mazouchi.
\newblock Optimal distributed learning for disturbance rejection in networked
  non-linear games under unknown dynamics.
\newblock \emph{IET Control Theory \& Applications}, 13\penalty0 (17):\penalty0
  2838--2848, 2018.

\bibitem[Tatari et~al.(2021)Tatari, Panayiotou, and
  Polycarpou]{Tatari2021FinitetimeIO}
Farzaneh Tatari, Christoforos Panayiotou, and Marios~M. Polycarpou.
\newblock Finite-time identification of unknown discrete-time nonlinear systems
  using concurrent learning.
\newblock \emph{2021 60th IEEE Conference on Decision and Control (CDC)}, pages
  2306--2311, 2021.

\bibitem[Thieme(2019)]{thieme2019multiflows}
Cameron Thieme.
\newblock Multiflows: A new technique for filippov systems and differential
  inclusions.
\newblock \emph{arXiv preprint arXiv:1905.07051}, 2019.

\bibitem[Tu et~al.(2013)Tu, Rowley, Luchtenburg, Brunton, and
  Kutz]{tu2013dynamic}
Jonathan~H Tu, Clarence~W Rowley, Dirk~M Luchtenburg, Steven~L Brunton, and
  J~Nathan Kutz.
\newblock On dynamic mode decomposition: Theory and applications.
\newblock \emph{arXiv preprint arXiv:1312.0041}, 2013.

\bibitem[Vahidi-Moghaddam et~al.(2020)Vahidi-Moghaddam, Mazouchi, and
  Modares]{vahidi2020memory}
Amin Vahidi-Moghaddam, Majid Mazouchi, and Hamidreza Modares.
\newblock Memory-augmented system identification with finite-time convergence.
\newblock \emph{IEEE Control Systems Letters}, 5\penalty0 (2):\penalty0
  571--576, 2020.

\bibitem[Vamvoudakis et~al.(2015)Vamvoudakis, Miranda, and
  Hespanha]{vamvoudakis2015asymptotically}
Kyriakos~G Vamvoudakis, Marcio~Fantini Miranda, and Jo{\~a}o~P Hespanha.
\newblock Asymptotically stable adaptive--optimal control algorithm with
  saturating actuators and relaxed persistence of excitation.
\newblock \emph{IEEE transactions on neural networks and learning systems},
  27\penalty0 (11):\penalty0 2386--2398, 2015.

\bibitem[Vapnik(2013)]{vapnik2013nature}
Vladimir Vapnik.
\newblock \emph{The nature of statistical learning theory}.
\newblock Springer science \& business media, 2013.

\bibitem[Walters et~al.(2018)Walters, Kamalapurkar, Voight, Schwartz, and
  Dixon]{walters2018online}
Patrick Walters, Rushikesh Kamalapurkar, Forrest Voight, Eric~M Schwartz, and
  Warren~E Dixon.
\newblock Online approximate optimal station keeping of a marine craft in the
  presence of an irrotational current.
\newblock \emph{IEEE Transactions on Robotics}, 34\penalty0 (2):\penalty0
  486--496, 2018.

\bibitem[Williams et~al.(2015)Williams, Kevrekidis, and
  Rowley]{williams2015data}
Matthew~O Williams, Ioannis~G Kevrekidis, and Clarence~W Rowley.
\newblock A data--driven approximation of the koopman operator: Extending
  dynamic mode decomposition.
\newblock \emph{Journal of Nonlinear Science}, 25\penalty0 (6):\penalty0
  1307--1346, 2015.

\bibitem[Williams et~al.(2016)Williams, Hemati, Dawson, Kevrekidis, and
  Rowley]{williams2016extending}
Matthew~O Williams, Maziar~S Hemati, Scott~TM Dawson, Ioannis~G Kevrekidis, and
  Clarence~W Rowley.
\newblock Extending data-driven koopman analysis to actuated systems.
\newblock \emph{IFAC-PapersOnLine}, 49\penalty0 (18):\penalty0 704--709, 2016.

\bibitem[Xia et~al.(2019)Xia, Zhang, Lu, and Zhou]{xia2019finite}
Yuanqing Xia, Jinhui Zhang, Kunfeng Lu, and Ning Zhou.
\newblock Finite-time tracking control of rigid spacecraft under actuator
  saturations and faults.
\newblock In \emph{Finite Time and Cooperative Control of Flight Vehicles},
  pages 141--169. Springer, 2019.

\bibitem[Yang and He(2019)]{yang2019adaptive}
Xiong Yang and Haibo He.
\newblock Adaptive critic learning and experience replay for decentralized
  event-triggered control of nonlinear interconnected systems.
\newblock \emph{IEEE Transactions on Systems, Man, and Cybernetics: Systems},
  50\penalty0 (11):\penalty0 4043--4055, 2019.

\bibitem[Yang et~al.(2014)Yang, Liu, and Wei]{yang2014near}
Xiong Yang, Derong Liu, and Qinglai Wei.
\newblock Near-optimal online control of uncertain nonlinear continuous-time
  systems based on concurrent learning.
\newblock In \emph{2014 International Joint Conference on Neural Networks
  (IJCNN)}, pages 231--238. IEEE, 2014.

\bibitem[Yang et~al.(2020)Yang, Vamvoudakis, Modares, Yin, and
  Wunsch]{yang2020safe}
Yongliang Yang, Kyriakos~G Vamvoudakis, Hamidreza Modares, Yixin Yin, and
  Donald~C Wunsch.
\newblock Safe intermittent reinforcement learning with static and dynamic
  event generators.
\newblock \emph{IEEE transactions on neural networks and learning systems},
  31\penalty0 (12):\penalty0 5441--5455, 2020.

\bibitem[Yeung et~al.(2019)Yeung, Kundu, and Hodas]{yeung2019learning}
Enoch Yeung, Soumya Kundu, and Nathan Hodas.
\newblock Learning deep neural network representations for koopman operators of
  nonlinear dynamical systems.
\newblock In \emph{2019 American Control Conference (ACC)}, pages 4832--4839.
  IEEE, 2019.

\bibitem[Zhang et~al.(2016)Zhang, Zhao, and Wang]{zhang2016event}
Qichao Zhang, Dongbin Zhao, and Ding Wang.
\newblock Event-based robust control for uncertain nonlinear systems using
  adaptive dynamic programming.
\newblock \emph{IEEE transactions on neural networks and learning systems},
  29\penalty0 (1):\penalty0 37--50, 2016.

\bibitem[Zhao et~al.(2015)Zhao, Zhang, Wang, and Zhu]{zhao2015experience}
Dongbin Zhao, Qichao Zhang, Ding Wang, and Yuanheng Zhu.
\newblock Experience replay for optimal control of nonzero-sum game systems
  with unknown dynamics.
\newblock \emph{IEEE transactions on cybernetics}, 46\penalty0 (3):\penalty0
  854--865, 2015.

\bibitem[Zhao and Liu(2020)]{zhao2020finite}
Zhijia Zhao and Zhijie Liu.
\newblock Finite-time convergence disturbance rejection control for a flexible
  timoshenko manipulator.
\newblock \emph{IEEE/CAA Journal of Automatica Sinica}, 8\penalty0
  (1):\penalty0 157--168, 2020.

\end{thebibliography}

%







\end{document}